\documentclass[3p]{elsarticle}

\usepackage{hyperref}
\usepackage{graphicx}
\usepackage{epstopdf, epsfig}
\usepackage{gensymb}
\usepackage{textcomp}
\usepackage{subfig}
\usepackage{amsmath}

\journal{International Journal of Heat and Fluid Flow}

\begin{document}

\begin{frontmatter}

\title{Enriching MRI mean flow data of inclined jets in crossflow with Large Eddy Simulations}

\author[mycorrespondingauthor]{Pedro M. Milani}
\cortext[mycorrespondingauthor]{Corresponding author}
\ead{pmmilani@stanford.edu}

\author{Ian E. Gunady, David S. Ching, Andrew J. Banko}
\author{\\Christopher J. Elkins, John K. Eaton}

\address{Mechanical Engineering Department, Stanford University \\
488 Escondido Mall, Stanford CA - 94305}

\begin{abstract}
Measurement techniques such as Magnetic Resonance Velocimety (MRV) and Magnetic Resonance Concentration (MRC) are useful for obtaining 3D time-averaged flow quantities in complex turbulent flows, but cannot measure turbulent correlations or near-wall data. In this work, we use highly resolved Large Eddy Simulations (LES) to complement the experiments and bypass those limitations. Coupling LES and magnetic resonance experimental techniques is especially advantageous in complex non-homogeneous flows because the 3D data allow for extensive validation, creating confidence that the simulation results portray a physically realistic flow. As such we can treat the simulation as data, which ``enrich'' the original MRI mean flow results. This approach is demonstrated using a cylindrical and inclined jet in crossflow with three distinct velocity ratios, $r=1$, $r=1.5$, and $r=2$. The numerical mesh is highly refined in order for the subgrid scale models to have negligible contribution, and a systematic, iterative procedure is described to set inlet conditions. The validation of the mean flow data shows excellent agreement between simulation and experiments, which creates confidence that the LES data can be used to enrich the experiments with near-wall results and turbulent statistics. We also discuss some mean flow features and how they vary with velocity ratio, including wall concentration, the counter rotating vortex pair, and the in-hole velocity.
\end{abstract}


\end{frontmatter}

\section{Introduction}

Magnetic Resonance Imaging (MRI) is a class of experimental techniques used across many scientific fields. In fluid mechanics, MRI-based methods can be used to measure 3D fields in complex turbulent engineering flows. The two main such methods are Magnetic Resonance Velocimetry (MRV), which is adapted for turbulent flows and measures three component mean velocity \citet{elkins20034dmagnetic}; and Magnetic Resonance Concentration (MRC), developed by \citet{benson2010three}, which is used to measure mean concentration of a passive scalar contaminant.

MRV and MRC are versatile techniques, with many advantages over more conventional diagnostics. First, they measure 3D fields in a Cartesian grid everywhere in the specified region of interest. Second, these techniques can be used in virtually any geometry, since they do not require optical access or the placement of probes. The working fluid must be water, and the test channel is limited in size (it must fit within an MRI scanner) and has to be made of non-magnetic materials. In practice, if a geometry can be 3D printed using stereolithography, the flow through it can be measured. Third, the relatively simple setup from conception to data acquisition allows for quick turnaround of experiments.

Several researchers have leveraged these capabilities to measure different complex turbulent flows. \citet{freudenhammer2014volumetric} used MRV to measure mean velocity in an internal combustion engine cylinder and presented detailed flow patterns around realistic intake valves. \citet{siegel2019design} measured the velocity field around a spinning projectile to study the Magnus force. \citet{shim20193d} applied MRV and MRC to a model urban canopy to understand how contaminants spread in a city under different wind conditions. \citet{ching2018investigation} used MRV measurements to study separated flows, and found significant geometric sensitivity of the size and secondary flows in the wake. Due to their flexibility, we expect that the number of groups employing these techniques will grow in the near future since MRI scanners are widely available in medical research facilities.

However, these MRI-based techniques have important limitations in engineering turbulent flows. Firstly, they measure mean velocity and concentration, but cannot directly measure other quantities such as pressure or turbulent statistics like the Reynolds stresses. Also, their resolution might be insufficient to capture some relevant small scale features of the mean turbulent flow. Finally, they do not provide reliable data close to solid boundaries due to inexact alignment and partial volume effects. These drawbacks can be quite significant in certain applications; for example, when designing improved turbulence models, the near-wall behavior is usually of particular interest and it would be crucial to know the turbulent statistics. 

The main contribution of the present paper is to demonstrate an approach for obtaining a detailed and reliable dataset whereby MRI data are acquired and subsequently enriched by highly resolved Large Eddy Simulations (LES). The idea is that the simulation geometry and inlet conditions should match the experiment as closely as possible, thus allowing the 3D MRI data to be used to thoroughly validate the numerical results. The simulation serves to complement the experiment and overcome the limitations inherent to MRI experiments. The advantage of performing MRI experiments alongside the Large Eddy Simulations instead of just running the LES is that the validation in 3D of the mean quantities provides confidence that the simulation results actually represent a physical flow and can be treated as data. In case of a mismatch, the 3D data helps to diagnose the root of the problem, which could be inlet conditions, averaging time, etc. Validation could be performed against point or plane data acquired with more traditional techniques such as thermocouple, hot-wire, or particle image velocimetry, but in complex turbulent flows those instruments could be more difficult to set up than the non-intrusive MRI. Besides, if low-dimensional experiments and high-fidelity simulation agree at sparse locations there is no guarantee that the simulation is actually capturing all the relevant 3-dimensional flow physics; if they do not agree, it is very difficult to uncover the causes. This is why the present approach is useful for generating trustworthy datasets in non-homogeneous turbulent flows. In the current work, this is demonstrated in inclined jets in crossflow.

\subsection{Inclined Jet in Crossflow}

\begin{figure}
  \begin{center}
  \includegraphics[width = 80mm] {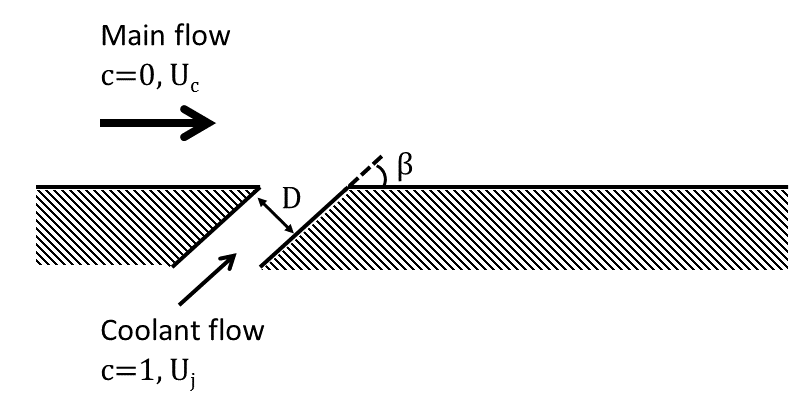}
  \end{center}
\caption{Circular inclined jet in crossflow, the geometry considered in the present paper. The jet contains a scalar contaminant whose dimensionless concentration is $c$.}
\label{fig-1-ijicf}
\end{figure}

The jet in crossflow is an important geometry in which fluid is ejected from an orifice and interacts with the flow passing above the orifice. The jet considered has a circular cross-section of diameter $D$ and is inclined with respect to the main flow by an angle $\beta$ as shown in Fig.~\ref{fig-1-ijicf}. For a review on the jet in crossflow literature, consult \citet{mahesh_review2013}. This geometry has several engineering applications. For example, gas turbine blades are cooled using film cooling, which roughly consists of inclined jets injecting cooler fluid into a hot crossflow to protect the solid walls from the high temperatures \citep{bogard_review}. Among the parameters that control the flow in a jet in crossflow in the incompressible regime is the velocity ratio $r$, defined as:
\begin{equation}
\label{eq-r}
r=U_j/U_c  
\end{equation}
where $U_j$ is the bulk velocity of the jet and $U_c$ is the bulk velocity of the crossflow. In general, low velocity ratio jets stay close to the wall, while high velocity ratio jets penetrate deep into the crossflow.

Previous studies have examined the jet in crossflow. \citet{fric1994vortical} used smoke visualization to describe different types of vortical structures present in a turbulent jet in crossflow. \citet{su2004simultaneous} and \citet{muppidi2008direct} studied scalar transport in a transverse ($\beta=90^\circ$) jet in crossflow with $r=5.7$, the former experimentally and the latter computationally. Among other things, they described decay rates and fluid entrainment. \citet{kohli2005turbulent} and \citet{schreivogel2016simultaneous} experimentally studied inclined jets at lower values of velocity ratio (which are more relevant for film cooling applications), and reported several turbulent mixing statistics. Some of their main conclusions pertain to the inadequacies of widely used turbulence models in this flow.

In the present paper, the same geometry is studied with three different values of velocity ratio, $r=1.0, 1.5, 2.0$. The MRV and MRC experiments are described in section 2 and the three LES's that are run to enrich the data are described in section 3. Section 4 discusses mean velocity and concentration results and validates the simulation, and then briefly presents LES turbulence data. Section 5 offers conclusions and paths for future work.

\section{Experimental setup and methods}

\subsection{Magnetic Resonance Imaging}

\begin{figure}
  \begin{center}
  \includegraphics[width = 80mm] {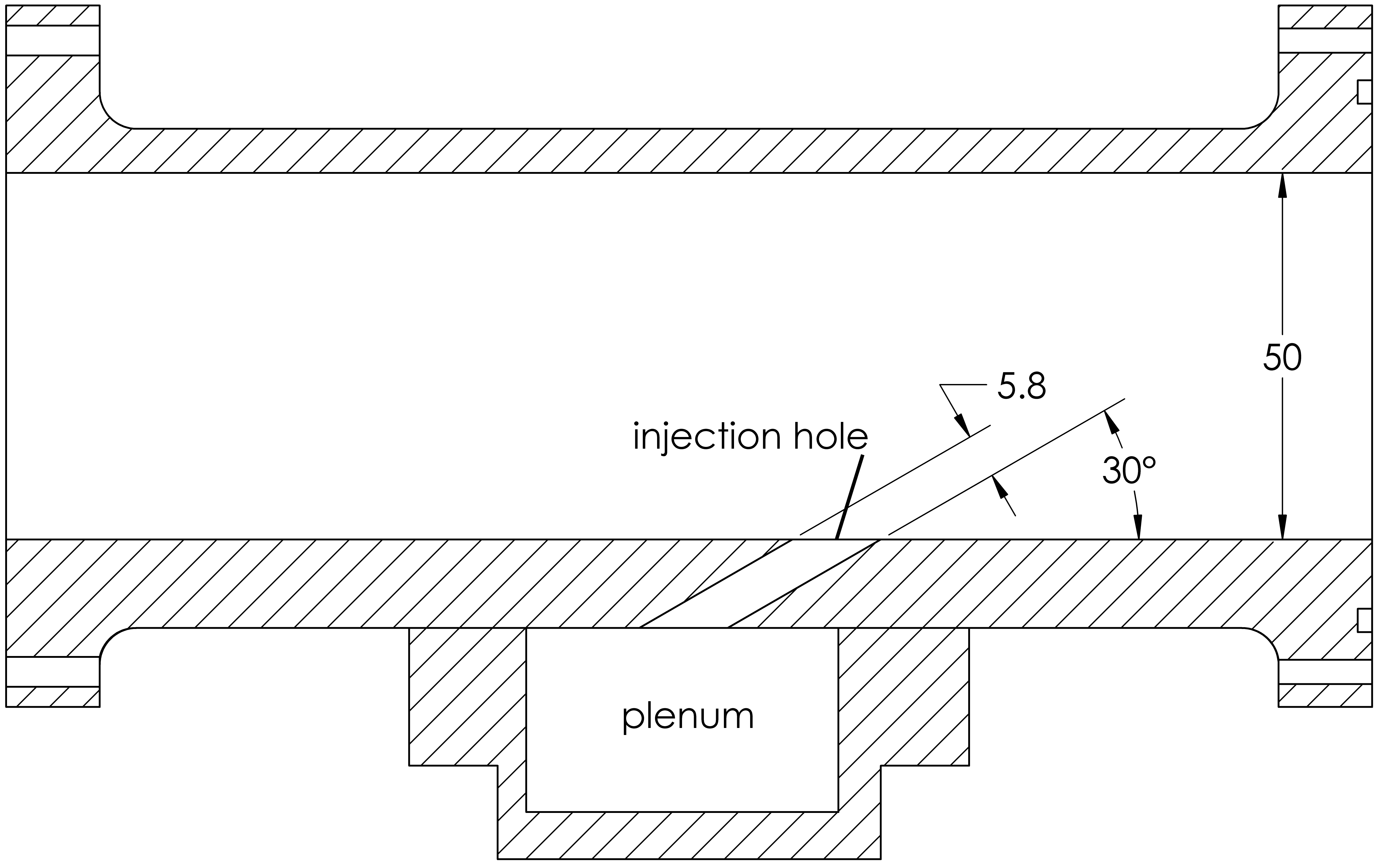}
  \end{center}
\caption{Schematic showing the test section, plenum, and injection hole. Dimensions in millimeters.}
\label{fig-2-schematicTest}
\end{figure}

Experiments are performed on an inclined jet in crossflow. The test section is shown in Fig.~\ref{fig-2-schematicTest}, and consists of a circular hole with diameter $D = 5.8$ mm, a non-dimensional length $L/D = 4.1$, and is inclined at $\beta = 30^\circ$. The main channel has a 50 mm by 50 mm square cross-section, and is large enough for the effects of confinement by the top and side walls to be negligible.

\begin{figure}
  \begin{center}
  \includegraphics[width = 160mm] {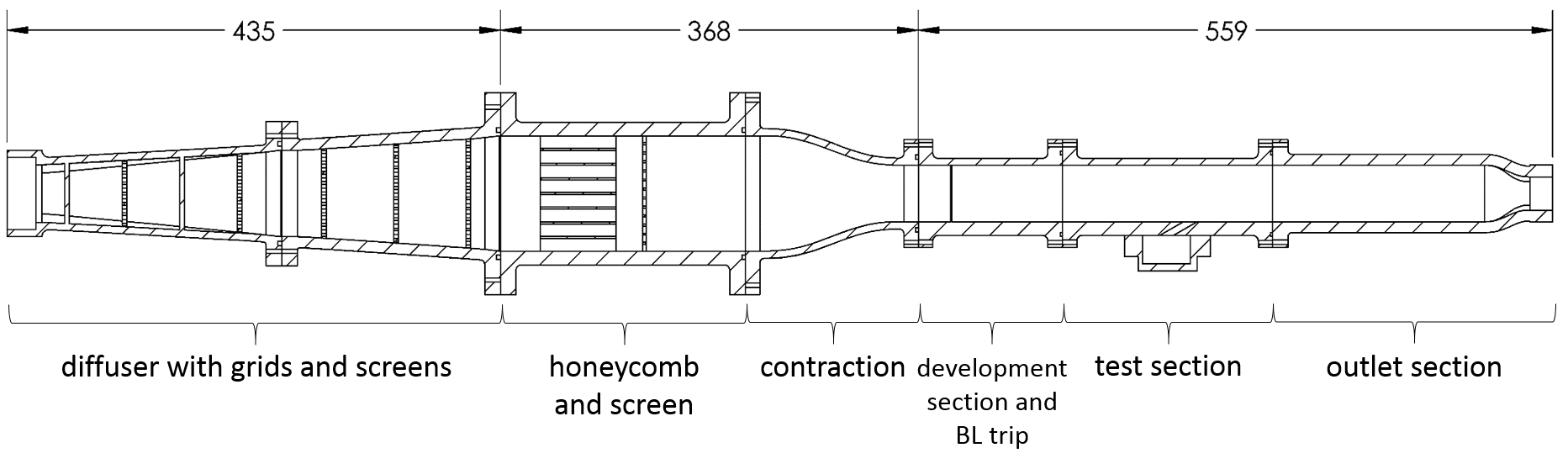}
  \end{center}
\caption{Channel schematic. Dimensions in millimeters. The inlet to the diffuser and plenum are connected to the flow meter, pumps, and reservoir with flexible tubing. The outlet returns flow to the reservoir through flexible tubing.}
\label{fig-3-schematicFull}
\end{figure}

A closed-loop flow circuit is used to provide the mainstream and injected flow, with water as the working fluid. The main channel flow is supplied by a 3/4 hp Little Giant pump (model TE-7-MD-SC) and the injected flow is supplied by a 1/8 hp Little Giant pump (model 5-MD-SC). Before reaching the 50 mm by 50 mm square test section, the main flow fluid is pumped from a reservoir and passes through a flow conditioning section consisting of a diffuser with screens to slow down the flow, a honeycomb section to minimize secondary flows, and a 4:1 contraction to minimize the boundary layer thickness and improve flow uniformity. A trip located on all four channel walls 210 mm upstream of the jet location initiates a new turbulent boundary layer. This geometry is shown in Fig. \ref{fig-3-schematicFull}. The injected flow is fed by a plenum attached to the test section. After injection, flow passes through the outlet of the test section and is returned to the reservoir, completing the closed-loop system.

During the experiment, the channel bulk velocity $U_c$ is maintained at 0.50 m/s and the jet bulk velocities $U_j$ are maintained at 0.5, 0.75, and 1.0 m/s for $r$ = 1.0, 1.5, and 2.0, respectively. Flow rates are monitored using Transonic PXL Flowsensors and are controlled using diaphragm valves. The reservoir temperature is held at approximately 21 \textdegree C throughout all experiments. The channel Reynolds number based on the test section width, channel bulk velocity, and kinematic viscosity of water at this temperature is $Re_C = 25,000$. The hole Reynolds number based on the hole diameter and jet bulk velocity varies from $Re_D = 2,900$ for $r = 1.0$ to $Re_D = 5,800$ for $r = 2.0$. The incoming boundary layer thickness and momentum thickness at the streamwise position of the hole center are measured using hot-wire anemometry (c.f. section 2.2) to be $\delta_{99} = 1.5D$ and $\theta = 0.15 D$, respectively. 

Magnetic Resonance Imaging (MRI) techniques are used to obtain the time-averaged, three-dimensional velocity and scalar concentration fields. Data are obtained on a volumetric Cartesian grid that includes the plenum, hole, and main channel over a streamwise length of about $10D$ upstream and $20D$ downstream of injection. The velocity and concentration data for $r=1.0$ were described in \citet{coletti2013turbulent}, with additional detail on methods and analysis in \citet{kevin_thesis}. Similar methods were used to collect the $r=1.5$ and $r=2.0$ cases in the present work. 

Velocity and concentration data were acquired using a 3 Tesla General Electric MRI scanner located at the Richard M. Lucas Center for Imaging at Stanford University. Magnetic Resonance Velocimetry (MRV) was used to obtain 3-component mean velocity data for both $r=1.5$ and $r=2.0$ on a Cartesian grid with a resolution of 0.66 mm by 0.6 mm by 0.6 mm in the streamwise, spanwise, and wall-normal directions, respectively. The MRV technique has been validated and described in detail in \citet{elkins20034dmagnetic}. Briefly, this technique measures the velocity of hydrogen protons in water. Copper sulfate is added at a concentration of 0.06 M to enhance the signal-to-noise ratio. Each experiment consists of 15 scans with the flow on and 6 scans with the flow off. Groups of three `on' scans are bookended by 'off` scans. Each such pair of `off' scans are averaged together and subtracted from the `on' scans. This procedure compensates for errors in velocity due to eddy-currents \citep{elkins20034dmagnetic}. Then, all 15 off-subtracted scans are averaged together. Finally, the mean velocity fields are filtered using the method described by \citet{schiavazzi2014matching} to reduce spurious noise and produce a nearly divergence free velocity field. During each individual scan, the MRV technique acquires data in spatial-Fourier space over a period of about 2-3 minutes. The entire procedure for a single blowing ratio takes approximately 50 minutes, which is significantly longer than all relevant flow time scales. 

Magnetic Resonance Concentration (MRC) was used to collect the mean scalar concentration field at $r=2.0$ on a Cartesian grid of the same resolution as the MRV measurements. The MRC technique has been previously described and validated in \citet{benson2010three}. Additional details of the technique, and modifications to flow reservoirs to mitigate contamination of the background signal by the scalar contaminant in a closed flow loop, are given in \citet{kevin_thesis}. Briefly, MRC uses a solution of copper sulfate as a contrast agent, where the measured signal magnitude is linearly proportional to the concentration. In a standard experiment, a reference concentration of 0.02 M is injected through the jet while the incoming mainstream flow is pure water (0 M). In practice, four scan types of 24 scans each were used to determine the concentration field. Reference and Background scans wherein the entire channel is filled with the reference concentration and pure water, respectively, were used to correct for large scale variations in the image magnitude due to the spatial sensitivity of the imaging coil. Then, standard scans with the reference concentration injected through the jet and a main flow of pure water are performed to image the mixing. Finally, inverted scans where pure water is injected through the jet and the main flow is held at the reference concentration were used to improve uncertainty in the near-wall data which can be affected by Gibbs artifacts \citep{benson2010three}. To construct the mean concentration field, the individual Standard and Inverted scans are background subtracted and normalized using the difference between the Reference and Background scans. The results are then averaged together to produce a concentration field which is unity in the plenum and zero in the freestream flow. Each individual scan acquires the spatial concentration field in Fourier space over a period of 2.5 minutes, and the entire experiment takes approximately 4 hours. Therefore, the mean concentration data are acquired over a time period which is orders of magnitude longer than all relevant flow time scales.

The uncertainty in mean velocity measurements due to thermal noise can be approximated from the following formula (\citet{pelc1994quantitative}):

\begin{equation}
\delta_{V} = \frac{\sqrt{2}}{\pi}\frac{V_{enc}}{SNR}
\label{eq_unc}
\end{equation}

Here $V_{enc}$ is the velocity encoding value which dictates the maximum measurable velocity. $SNR$ is the signal to noise ratio defined as the mean signal magnitude in a region of interest downstream of injection divided by the root-mean-square signal within solid regions. For $r=1.5$ and $r=2.0$, the value of $V_{enc}$ in the streamwise, wall-normal, and spanwise directions were 1.5 m/s, 1.4 m/s, and 1.0 m/s, respectively, and the $SNR$ was approximately 35. This gives the average uncertainty across each velocity component as 5$\%$ of the main flow bulk velocity. In practice, the velocity uncertainty is due to signal loss from turbulent dephasing, thermal noise, and statistical uncertainty. However, it has been found that this formula is in good agreement with statistics-based estimates determined from variations between scans and with the typical deviation of MRV from techniques such as hot-wire and PIV in validation test cases (\citet{pelc1994quantitative}). 

The uncertainty in the mean concentration measurements was determined from a statistical procedure based on the variations between individual Reference, Background, Standard, and Inverted scans. The uncertainty is then propagated through the equation used to compute the concentration field. The details of this procedure are given in \citet{kevin_thesis}. The uncertainty is analyzed in a region of interest in the mixing region of the main channel, defined as the region downstream of injection where the concentration is greater than 5$\%$. The uncertainty in mean concentration is 4$\%$ and 6$\%$ of the injected concentration for $r=1.0$ and $r=2.0$, respectively.

The review of \citet{elkins2007review} contains more information on the technique and possible use cases of Magnetic Resonance Velocimetry. One important limitation of the present study when compared to realistic jet in crossflow applications is that the working fluid is a liquid and there is no density difference between jet and crossflow. It is possible to mitigate either of those, though that could significantly increase the complexity of the experimental setup and thus is rarely done. For example, some workers used fluorinated gases as a working fluid in MRI velocimetry instead of water (\citet{elkins2007review}). It is also possible to vary the properties of the cooling jet within a small band by mixing different liquids with water, such as alcohol or glycerin. Depending on the physics under consideration, however, those extra steps might not be necessary. For instance, \citet{yapa2014comparison} compared MRC measurements in water with temperature measurements in air flow with Mach number of 0.7 in the same plane shear layer apparatus and obtained good agreement, demonstrating that the water channel utilizing a temperature-scalar analogy was sufficient to capture the relevant phenomena present in a subsonic compressible flow.

\subsection{Hot-wire anemometry}

A hot-wire experiment was performed to better characterize the inlet profiles of the main channel flow and to inform inlet condition specification for numerical simulations, as described in section 3.4. The same channel that was used in the water-based MRI experiments (Fig. \ref{fig-3-schematicFull}) was run in air with the same Reynolds number, corresponding to a bulk velocity of $U_c = 7.5$ m/s at a temperature of $23 ^\circ C$. The outlet section of the channel was removed, flow was only present in the main channel, and the injection hole was blocked off, flush to the bottom wall. A hot-wire probe was placed in the symmetry plane of the channel ($z/D=0$), at the same streamwise position as the center of the injection hole ($x/D=0$), and its distance from the bottom wall was carefully controlled using a traverse driven by a stepper motor. The probe is 3 mm long, and a 5 $\mu$m diameter gold-plated wire is mounted, with an active section that is 1 mm long. The overheat ratio employed is 1.8, and the calibration described in \citet{hultmark2010temperature} is used, including the temperature correction. A conservative estimate for the calibration uncertainty puts the uncertainty of the velocity measurements at $1.5\%$ when the velocity is relatively low ($u/U_c = 0.5$), and at $0.5\%$ when the velocity is relatively high ($u/U_c = 1.5$).

\section{Computational setup}

Three distinct Large Eddy Simulations were conducted, one for each of the velocity ratios ($r=1.0, 1.5, 2.0$). In the current work, the high-fidelity simulations serve to enrich the 3D mean field from MRV/MRC with turbulent statistics and time-resolved quantities. For that, it is crucial to simulate the same flow conditions used in the experiment and thoroughly validate the LES results. Fine meshes are employed to ensure that the viscous sublayer is adequately resolved and that the subgrid scale model contributions are negligible in interesting regions of the flow. More details on the simulations are presented in the following subsections.

\subsection{Governing equations}

The filtered, incompressible continuity and Navier-Stokes equations are solved in 3D as shown in Eq.~\ref{eq-continuity} and Eq.~\ref{eq-ns}.

\begin{equation}
\label{eq-continuity}
\frac{\partial \tilde{u_k}}{\partial x_k} = 0 \newline
\end{equation}
\begin{equation}
\label{eq-ns}
\frac{\partial \tilde{u}_i}{\partial t} + \frac{\partial \left( \tilde{u}_j \tilde{u}_i \right)}{\partial x_j} = -\frac{1}{\rho} \frac{\partial \tilde{p}}{\partial x_i} + \nu \frac{\partial^2 \tilde{u}_i}{\partial x_j \partial x_j} - \frac{\partial}{\partial x_j} \tilde{\tau}_{ij}
\end{equation}

The tilde over a variable is shown explicitly to denote grid-filtered quantities; $u_i$ are the Cartesian components of the velocity, and $p$ is the pressure. The unresolved scales are accounted for with the subgrid scale stress, $\tilde{\tau}_{ij} = \widetilde{u_i u_j} - \tilde{u}_i \tilde{u}_j$. The fluid properties, density $\rho$ and kinematic viscosity $\nu$, are constant.

The filtered equation for a passive scalar $c$ is also solved, as shown in Eq.~\ref{eq-scalar}. The molecular Schmidt number is given by $Sc$ and the subgrid scale mixing is represented by $\tilde{\sigma}_{j} = \widetilde{c u_j} - \tilde{c} \tilde{u}_j$.

\begin{equation}
\label{eq-scalar}
\frac{\partial \tilde{c}}{\partial t} + \frac{\partial \left( \tilde{u}_j \tilde{c} \right)}{\partial x_j} = \frac{\nu}{Sc} \frac{\partial^2 \tilde{c}}{\partial x_j \partial x_j} - \frac{\partial}{\partial x_j} \tilde{\sigma}_{j}
\end{equation}

To match the properties of water, used in the experiments, the values of density and viscosity were set to $\rho=998 kg/m^3$ and $\nu=1.0 \times 10^{-6} m^2/s$. Thus, the Reynolds number based on hole diameter $D$ and bulk jet velocity $U_j$ varies between $Re_D = 2900$ and $Re_D=5,800$. The value of the molecular Schmidt number $Sc$ was set to $1.0$ in the simulations to ensure that the smallest length scales of the velocity and scalar fields match and are both resolved adequately. This is much smaller than the true molecular Schmidt number of copper sulfate in water, which is of the order of $10^3$ \citep{emanuel1963diffusion}. However, this discrepancy does not have any practical effect on the scalar field computed: in both cases ($Sc=1$ or $Sc=10^3$), molecular diffusion is negligible compared to turbulent mixing in this flow. Preliminary RANS simulations confirmed that the mean scalar field is insensitive to $Sc$ in this range except extremely close to the bottom wall, where molecular diffusion might compete with turbulent mixing. This coincides with the region where MRC data are unreliable, so there are no scalar data to be matched there. Furthermore, since zero scalar flux boundary condition is maintained at all walls, diffusion is insignificant in that region.

A subgrid scale model is required to determine $\tilde{\tau}_{ij}$ and $\tilde{\sigma}_{j}$ in equations \ref{eq-ns} and \ref{eq-scalar}. For the momentum equation, the Vreman model is employed \citep{vreman2004eddy}. This is a linear eddy viscosity model where the eddy viscosity $\nu_{t,SGS}$ depends on the local velocity gradients and cell spacing. For the scalar equation, gradient diffusion and the Reynolds analogy are employed, with turbulent diffusivity $\alpha_{t,SGS} = \nu_{t,SGS}/Sc_{t,SGS}$. The subgrid scale turbulent Schmidt number was set to a constant, $Sc_{t,SGS}=0.85$.

Henceforth, the tilde will be omitted when referring to filtered variables to simplify notation. For example, when referring to the LES, $u_i$ will denote the result of the simulation (which is the filtered quantity), $\bar{u}_i$ will represent the time average of the filtered quantity, and primes will be used to refer to fluctuating quantities in the filtered sense (i.e. $u'_i \equiv u_i - \bar{u}_i$). 

\subsection{Solver}

The equations of section 3.1 are solved using the unstructured and incompressible solver Vida, developed by Cascade Technologies. It employs the method developed by \citet{ham2007efficient} which consists of second-order accurate spatial discretization and explicit time advancement, based on second-order Adams Bashforth. Parallelization is implemented using MPI and each computation is performed on up to 1024 cores. The total computational cost of each simulation varied between $5.2 \times 10^5$ and $1.2 \times 10^6$ CPU-hours as shown in Tab.~\ref{tab-1-cases}.

\subsection{Domain and mesh}

\begin{figure}
    
  \begin{center}
  \includegraphics[width = 160mm] {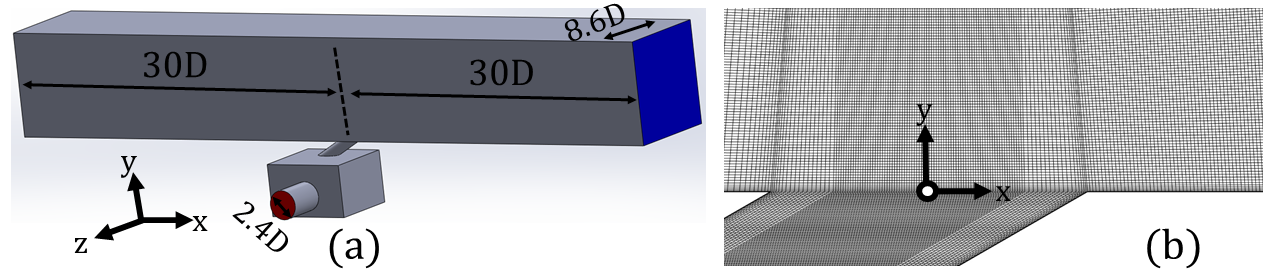}
  \end{center}
  
\caption{(\textit{a}) Simulation domain, where walls are shown in grey, coolant inlet is shown in red, and outlet is shown in blue. (\textit{b}) Mesh in the centerplane, around the region where the circular hole meets the bottom wall; the axis is centered on the origin.}
\label{fig-4-mesh}
\end{figure}

The full domain simulated in the LES is shown in Fig.~\ref{fig-4-mesh}(a). The crossflow is aligned with the $+x$ direction and the jet comes from a circular hole inclined $30^\circ$ with respect to the streamwise direction. The main channel inlet is located $30D$ upstream of the center of the hole exit, where the origin is located, and extends up to $x/D = 30$. In the experiment, the coolant flow is fed from a rectangular plenum located under the channel. The simulation domain includes the plenum and part of the tube that feeds it following best practices of the film cooling community \citep{walters1997systematic}. The plenum feed is $3.4D$ long and has a circular cross-section of diameter $2.4D$. The plenum inlet is marked in red, and the main channel outlet is marked in blue.

The mesh is block-structured and contains only hexahedral elements. It was generated with the software ANSYS ICEM. We aimed to have adequate resolution along the bottom wall and the jet hole wall, so that no wall models would be needed in those critical regions. The spacing of the first cell above the wall was set based on preliminary RANS simulations, and the LES results confirm that in all cases the first cell is located at $y^+ < 1.5$ for the bottom wall and at $y^+ < 3.0$ for the cooling hole (with a large majority of cells located at $y^+ < 1.0$). We also required fine overall spacing inside the hole and in the region where the jet and crossflow interact; a typical mesh element there measures approximately $0.02D$ in each Cartesian direction. Fig.~\ref{fig-4-mesh}(b) shows the mesh, and Tab.~\ref{tab-1-cases} contains the total number of mesh points utilized in each of the three simulations.

\begin{table}
  \begin{center}
  \begin{tabular}{|l | c | c | c |}
       \hline
       $r$  & Number of cells & Time step [ $D/U_c$ ] & Computational cost [CPU-hours] \\[3pt]
       \hline
       $1.0$ & $40.1M$ & $5.0 \times 10^{-3}$ & $5.2 \times 10^5$\\
       $1.5$ & $43.4M$ & $4.1 \times 10^{-3}$ & $9.8 \times 10^5$\\
       $2.0$ & $48.3M$ & $3.4 \times 10^{-3}$ & $1.2 \times 10^6$\\
       \hline
  \end{tabular}
  \caption{Total mesh size, dimensionless time step used, and total computational cost for the LES of each velocity ratio.}
  \label{tab-1-cases}
  \end{center}
\end{table}

\begin{figure}

  \begin{center}
    \subfloat[]{{\includegraphics[width=55mm]{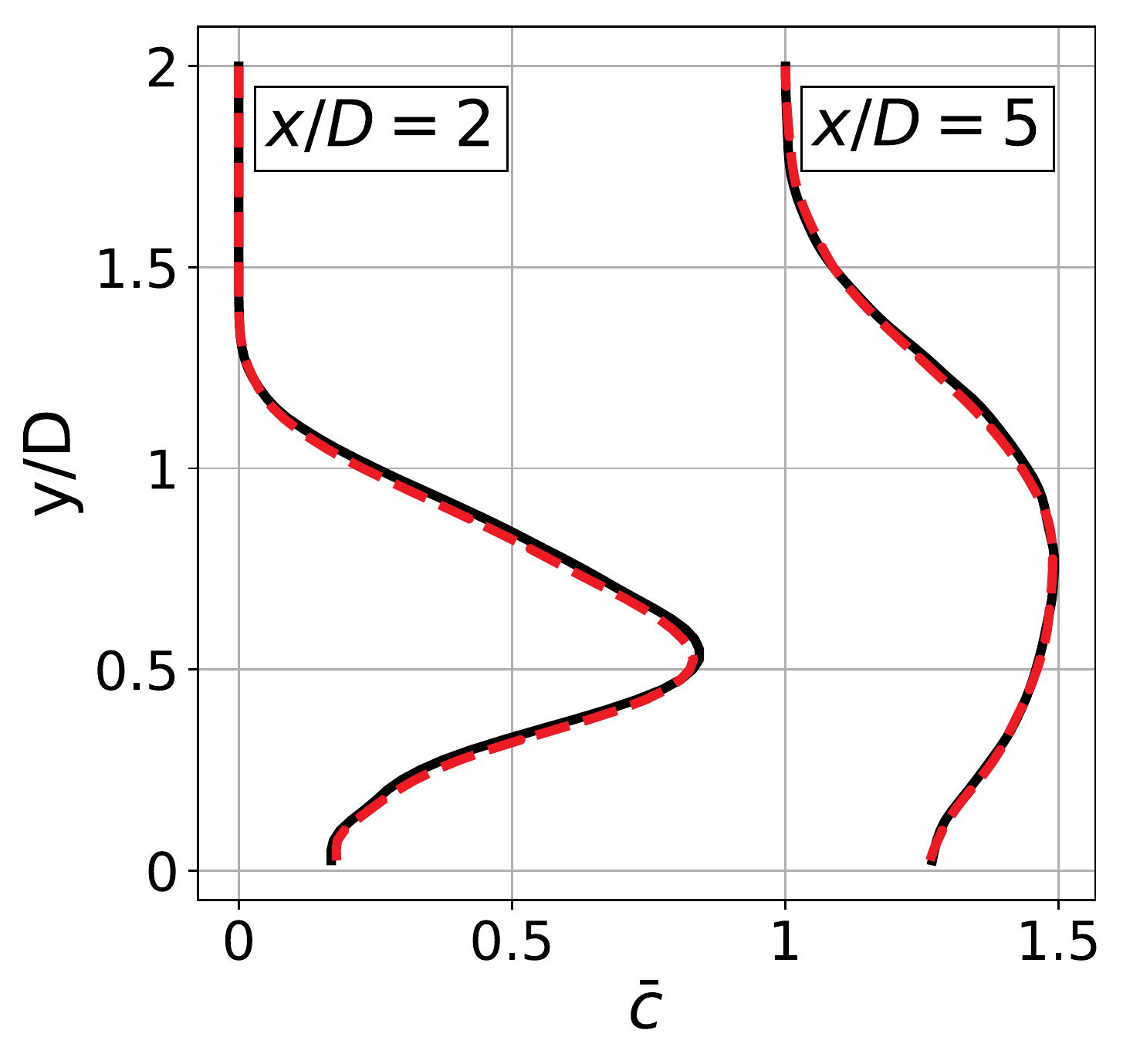} }}%
    \subfloat[]{{\includegraphics[width=54mm]{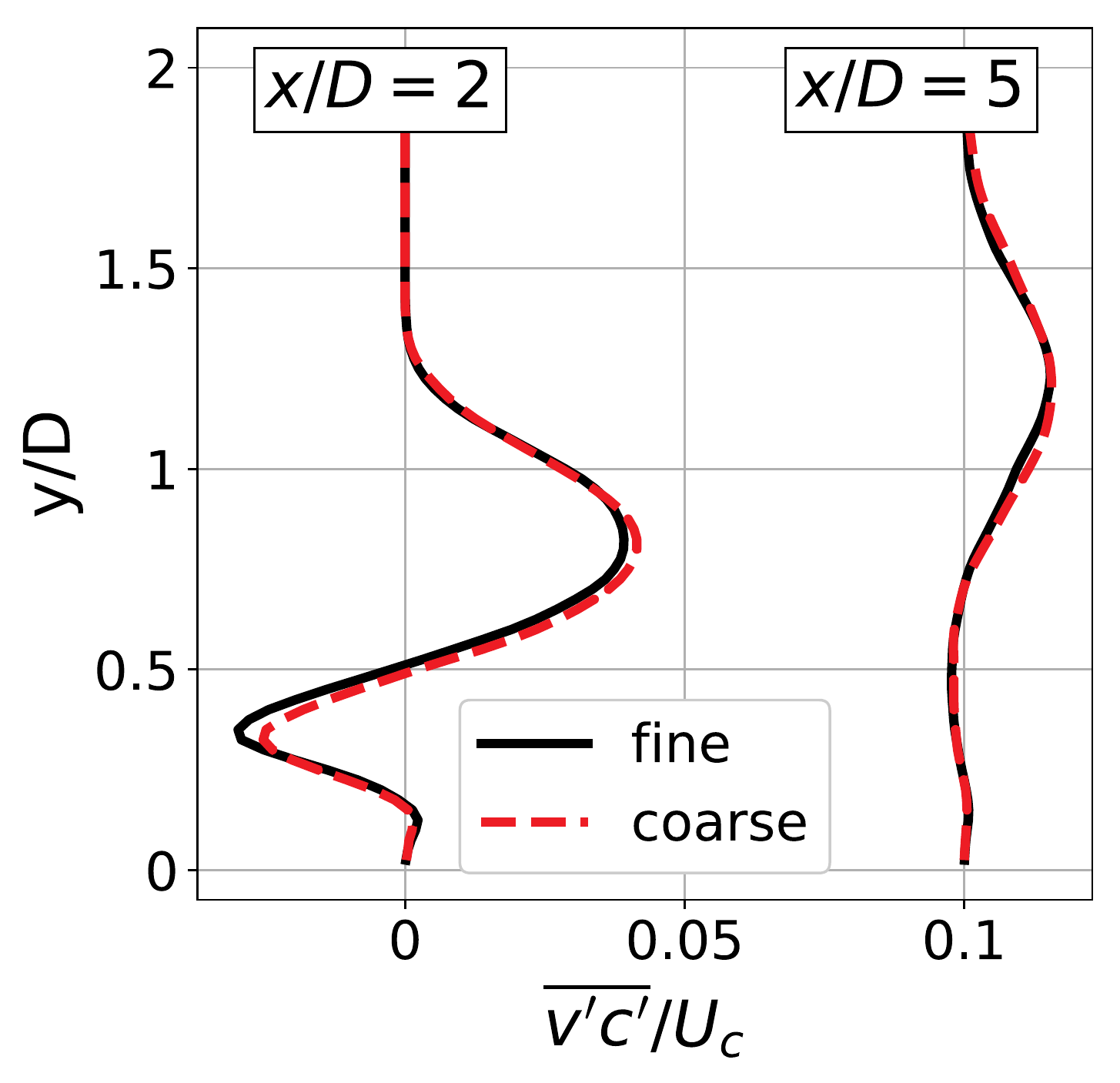} }}
  \end{center}
  
\caption{Plots showing vertical mean profiles from the coarse and fine $r=1$ LES at $z/D=0$. (\textit{a}) shows mean scalar $\bar{c}$ and (\textit{b}) shows vertical turbulent scalar flux $\overline{v'c'}$. Two streamwise locations are shown; for $x/D=5$, the lines are shifted to the right by $1.0$ in (\textit{a}) and $0.1$ in (\textit{b}).}
\label{fig-5-meshconv}
\end{figure}

\begin{figure}
    
  \begin{center}
  \includegraphics[width = 60mm] {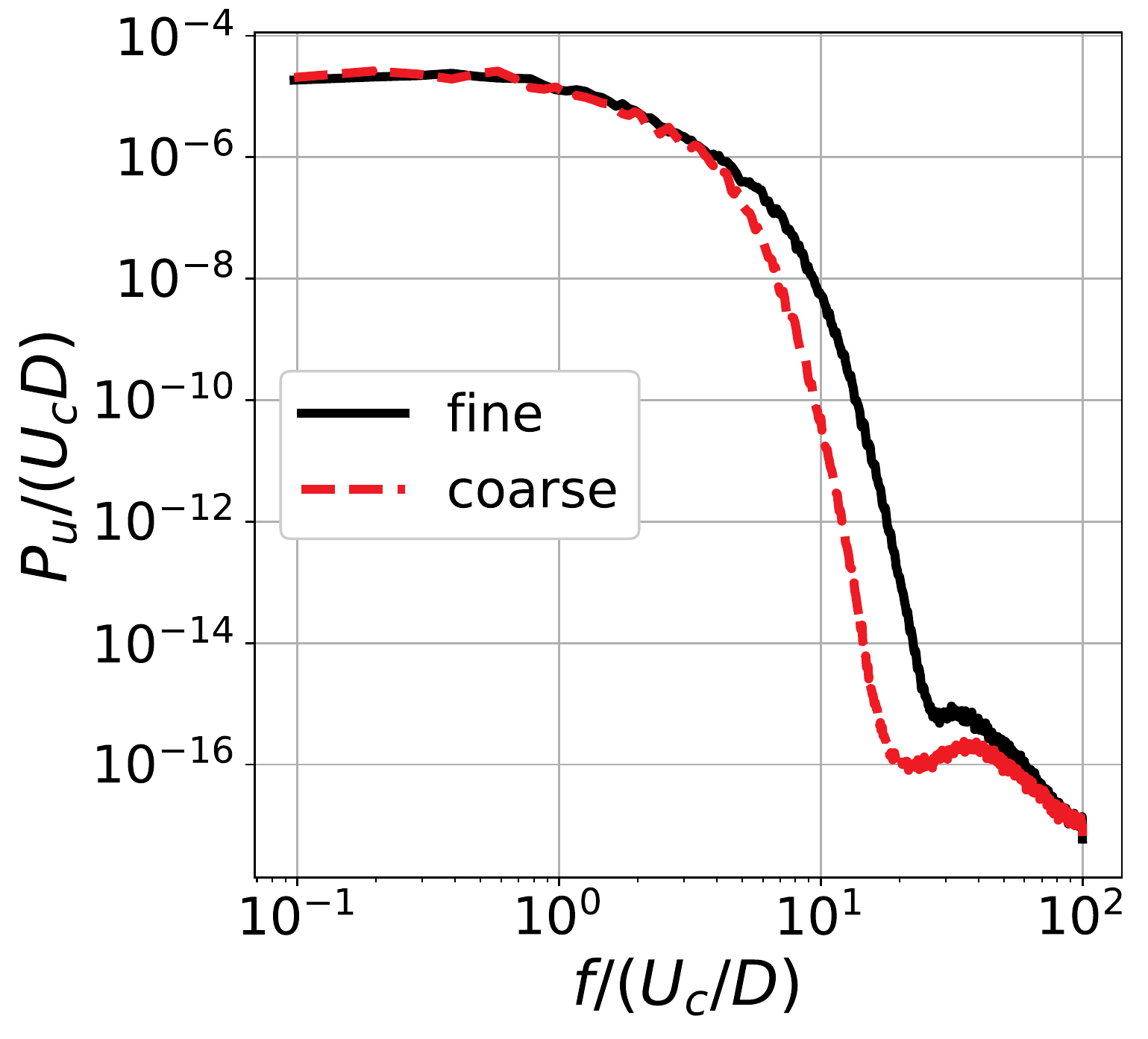}
  \end{center}
  
\caption{Log-log plot comparing the power spectrum density of $u$ versus frequency $f$ in the $r=1$ case. Time sequence of $u$ is probed at $x/D=3$, $y/D=1$ and $z/D=0$ (marked as point C in Fig.~\ref{fig-7-convergencelocations})}
\label{fig-6-meshspectra}
\end{figure}

To verify that the mesh convergence is satisfactory, a coarser simulation was run for the $r=1$ case. The mesh in the interesting regions of the flow was made coarser by a factor of approximately $1.6$ in all three directions, yielding a mesh with $16.8M$ cells instead of $40.1M$. The coarse simulation ran with the same time step and for the same total time. Fig.~\ref{fig-5-meshconv} contains comparisons between the coarse and fine LES. Vertical profiles of $\bar{c}$ and $\overline{v'c'}$ in two different streamwise locations show excellent agreement between coarse and fine meshes, which suggests good mesh convergence. The power spectral density $P_u$ was also calculated for a time sequence of the streamwise velocity $u$ at a probe located in the windward shear layer ($x/D=3$, $y/D=1$, $z/D=0$). The plot of non-dimensional $P_u$ as a function of the frequency $f$ is shown in Fig.~\ref{fig-6-meshspectra}. As expected, it shows that the coarse simulation captures less of the high-frequency variation. However, it is already able to resolve $93\%$ of the variance resolved by the fine simulation, which again suggests that the fine mesh has an appropriate resolution to capture almost all relevant scales of this flow. All subsequent results reported in this paper pertain to the fine mesh LES.

Another important metric to evaluate the mesh resolution is the subgrid scale viscosity, $\nu_{t,SGS}$. For all three simulations, the mean subgrid scale viscosity is less than half of the molecular viscosity everywhere where jet and crossflow interact. This means that the modeled turbulent fluctuations have a negligible impact on the mean flow. Together with the mesh refinement study, this shows that the fine mesh simulations, despite being formally Large Eddy Simulations, have mesh resolutions that are close to that required of Direct Numerical Simulations (DNS).

\subsection{Time step and averaging}

\begin{figure}
  \begin{center}
  \includegraphics[width = 75mm] {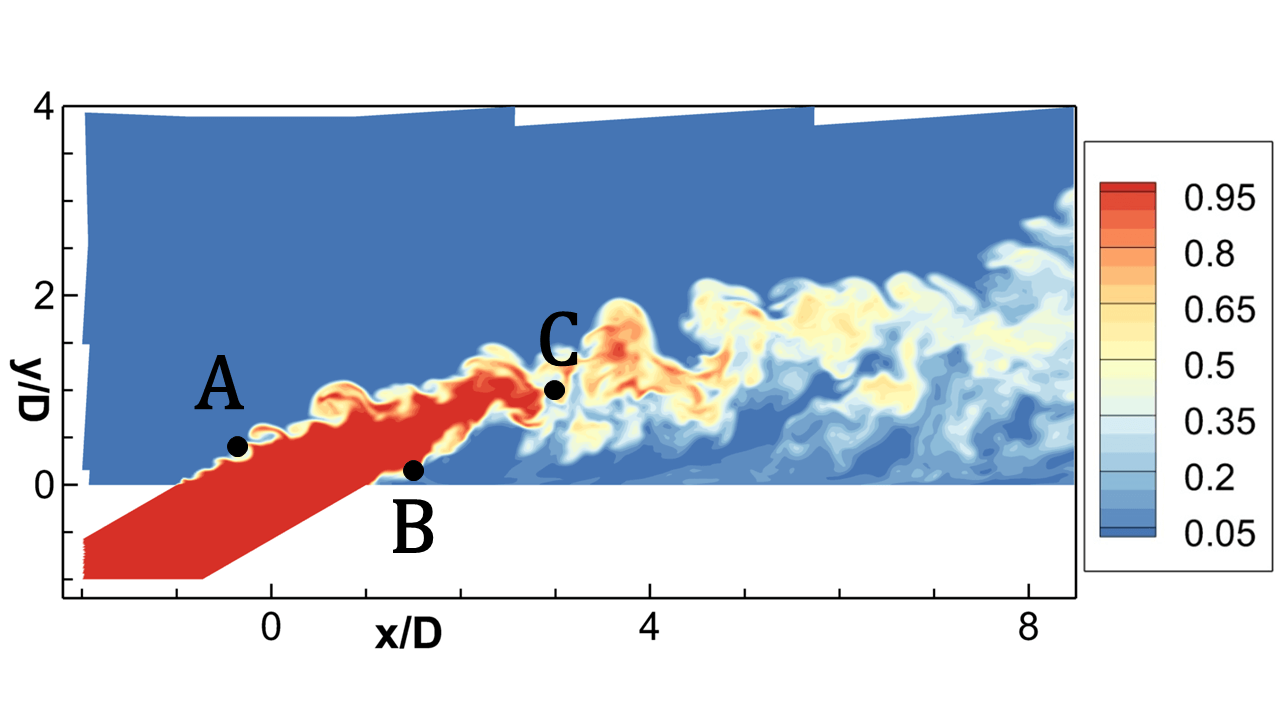}
  \end{center}
\caption{Instantaneous snapshot of the $r=2$ simulation, with contours of the scalar contaminant $c$. This plot shows locations A, B, and C whose time-resolved behavior is analyzed. All points are located on the symmetry plane, $z/D=0$. A is located at $x/D=-0.4$ and $y/D=0.4$; B is located at $x/D=1.5$ and $y/D=0.17$; C is located at $x/D=3$ and $y/D=1$.}
\label{fig-7-convergencelocations}
\end{figure}

\begin{figure}

  \begin{center}
    \subfloat[]{{\includegraphics[width=55mm]{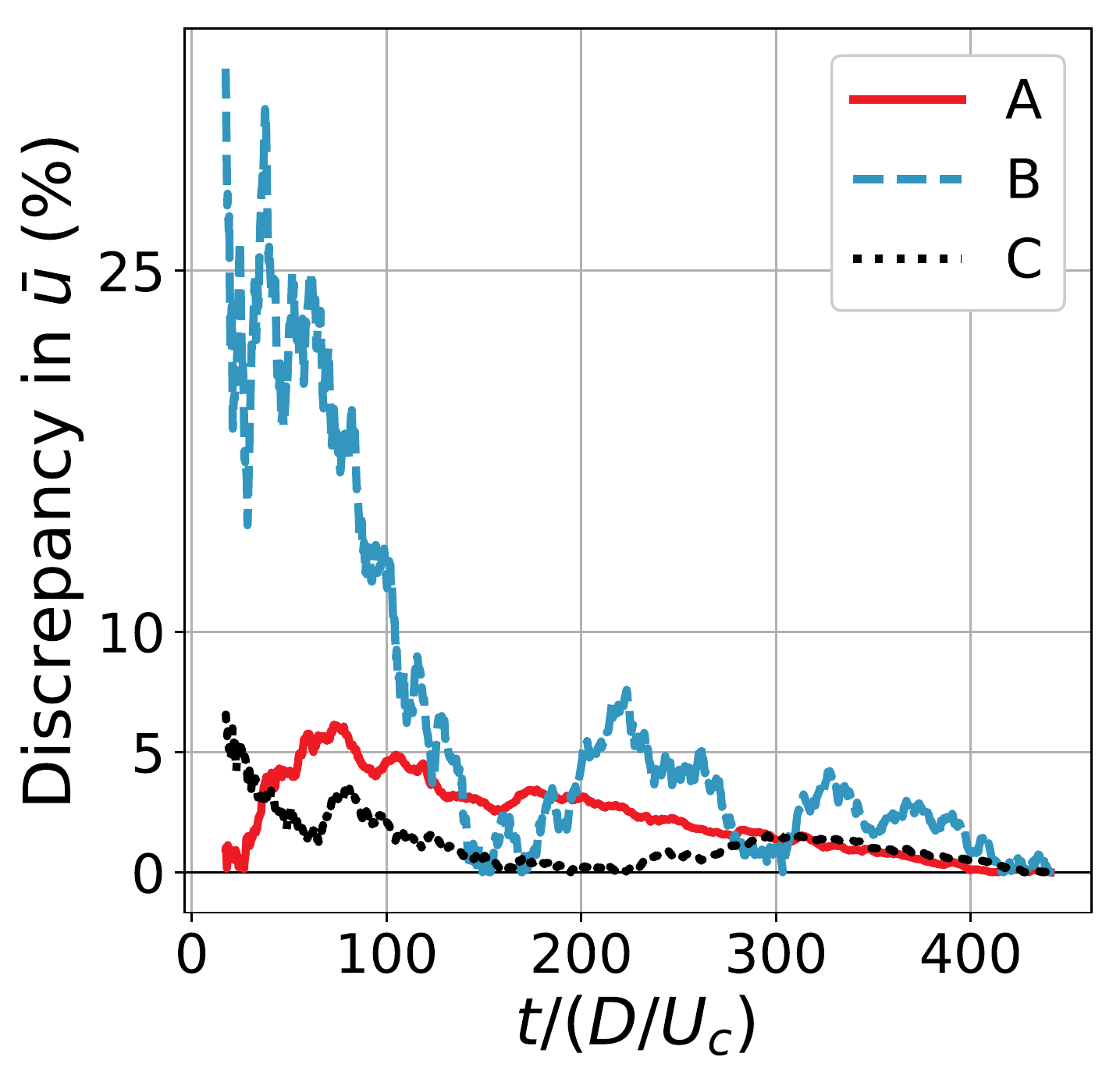}}}%
    \subfloat[]{{\includegraphics[width=55mm]{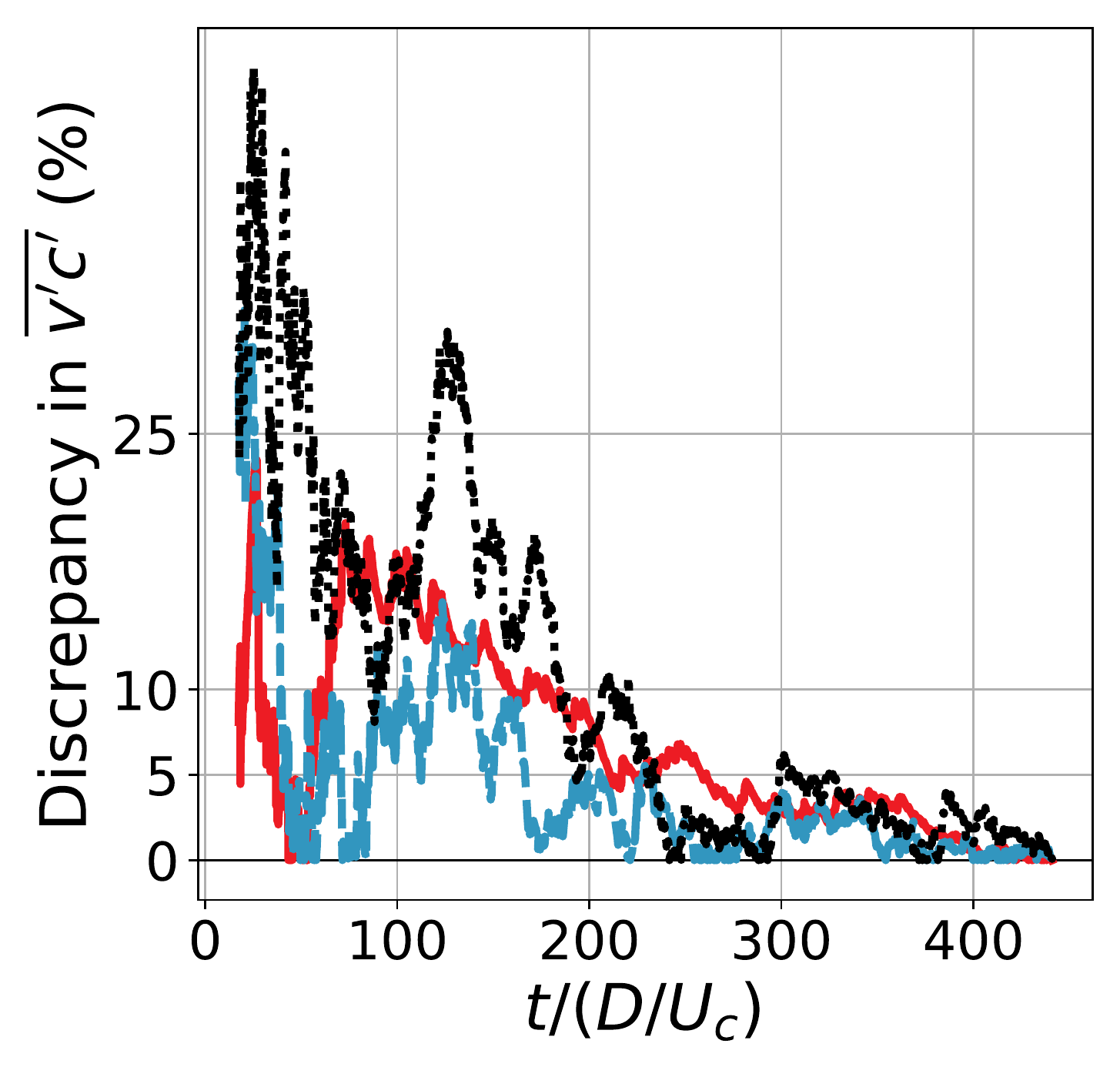}}}%
  \end{center}
  
\caption{Percent difference between perceived time averages at each time step and the time average at the end of the simulation in the $r=2$ case. The three spacial locations (A,B,C) are detailed in Fig.~\ref{fig-7-convergencelocations}}
\label{fig-8-convergence}
\end{figure}

The time step for the Large Eddy Simulations was chosen based on the Courant-Friedrichs-Lewy (CFL) number calculated with the crossflow bulk velocity, $U_c$, and the smallest mesh element anywhere in the domain, which is that immediately above the hole wall. We set the CFL number to $1.0$, and report the resulting time steps in Tab.~\ref{tab-1-cases}. Note that higher values of $r$ require smaller time steps (which scale as $r^{-0.5}$), since the mesh spacing must be reduced next to the hole wall by the same factor to maintain similar $y^+$.

All simulations are initialized with $u=U_c$, $v=w=0$ everywhere, and $c=1$ in the plenum and in the cooling hole and $c=0$ in the main channel. They were run for $240 D/U_c$ to achieve a statistically stationary state; thereafter, statistics were collected every time step for a simulation time of at least $440 D/U_c$, as specified in in Tab.~\ref{tab-1-cases}. These simulation times are deemed sufficient for convergence through a comparison of statistics obtained with increasing simulation time. Furthermore, they are significantly longer than the simulation times used by previous authors who computed similar flows \citep[e.g.][]{muppidi2007direct, bodart2013highfidelity}. 

To estimate uncertainty due to incomplete time averaging, we track mean velocity and second order turbulence statistics at three different locations of the flow, shown in Fig.~\ref{fig-7-convergencelocations}. At each time step $t$, we calculate the time average if the simulation had only run up until that instant. The discrepancy as a function of simulation time, then, is defined as the absolute percent difference between the time average at that time step $t$ and the time average at the final time step (which is the average ultimately reported). Fig.~\ref{fig-8-convergence} shows plots of such discrepancy for the $r=2$ case and for two different mean quantities, the mean streamwise velocity $\bar{u}$ and mean vertical turbulent scalar flux, $\overline{v'c'}$. The plots show that the oscillations in the perceived statistics are initially high, but as the simulation progresses the oscillations decay indicating lower uncertainty in the statistics. Towards the end of the simulation, such oscillations indicate that our uncertainty in both quantities is well under $5\%$.

It is important to note that the streamwise velocity averages converge much more rapidly than the higher order correlations. A consequence is that it is important to average Large Eddy Simulations for a relatively long time in such fully non-homogeneous geometries if one desires to use the resulting dataset to quantitatively analyze turbulent correlations such as $\overline{u_i'c'}$. For example, \citet{bodart2013highfidelity} averaged their similar simulation for $200 D/U_c$; in Fig.~\ref{fig-8-convergence}(b), it is clear that the values of $\overline{v'c'}$ can vary by more than $10\%$ between $t=200 D/U_c$ and the end of the present simulation, at $t=440 D/U_c$. This suggests that a simulation time of $200 D/U_c$ might lead to insufficiently averaged statistics.

\subsection{Inlet and boundary conditions}

\begin{table}
        
        \hrule
        \renewcommand{\labelenumi}{\arabic{enumi}}
        
        \begin{enumerate}
            \item Guess initial parameters for the inlet condition generator.
            \item Run a channel flow LES with given inlet condition until convergence.
            \item Compare $\bar{u}(y)$ and $u'_{rms}(y)$ between the hot-wire data and different streamwise stations of the channel flow LES:
                \begin{enumerate}
                    \item If none of the profiles at different stations agree with the hot-wire data, modify inlet conditions and repeat (2)-(3).
                    \item If station located at distance $L_{station}$ from the inlet of the channel LES matches the hot-wire data well, we are done. For the jet in crossflow simulation, we use the current inlet condition and set the main inlet to be located a distance $L_{station}$ upstream from the hole. 
                \end{enumerate}
        \end{enumerate}
    
    \hrule
        
    \caption{Iterative procedure to determine inlet conditions}
    \label{tab-2-inletalgorithm}
\end{table}

\begin{figure}

  \begin{center}
    \subfloat[]{{\includegraphics[width=52mm]{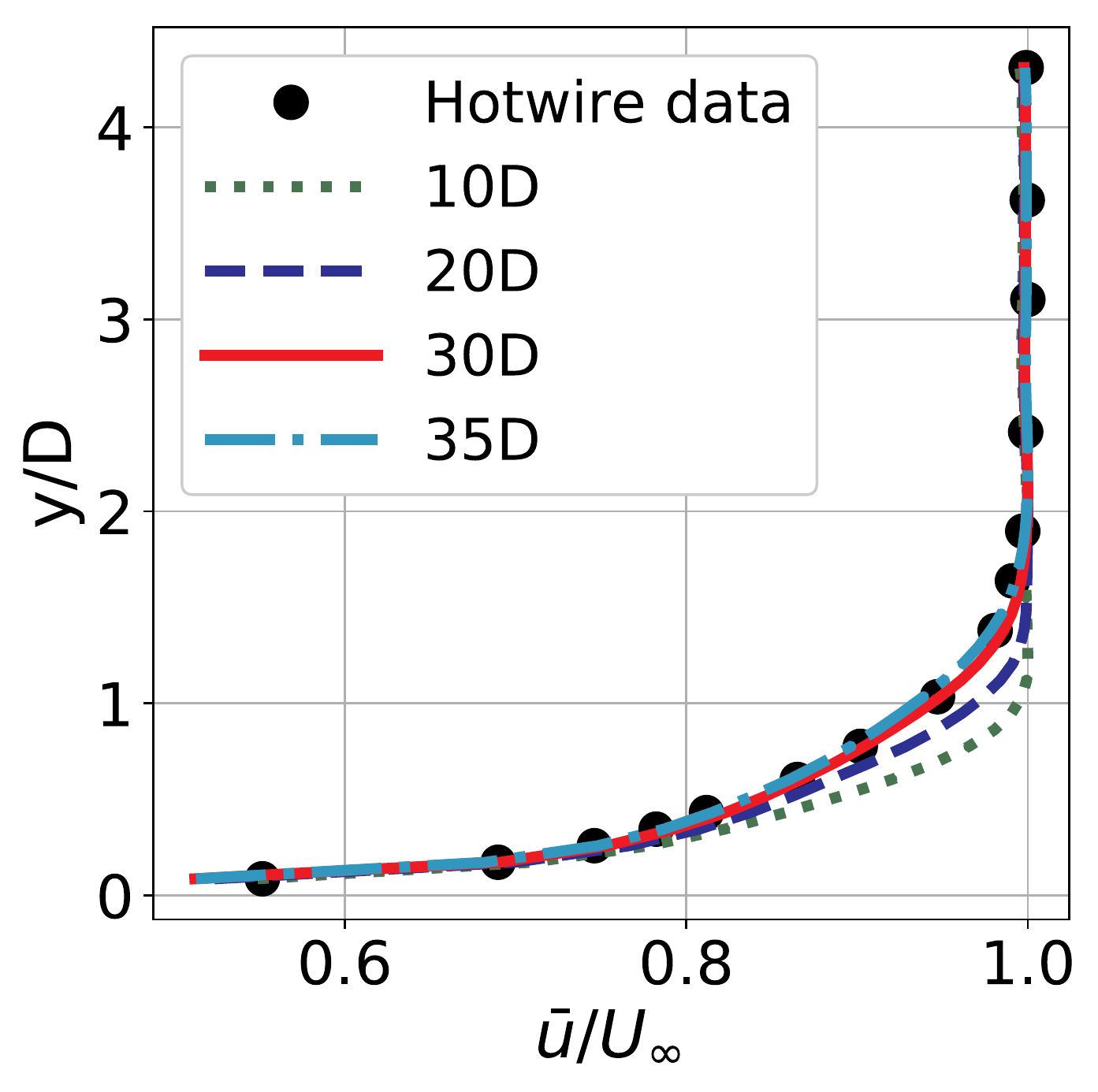}}}%
    \subfloat[]{{\includegraphics[width=52mm]{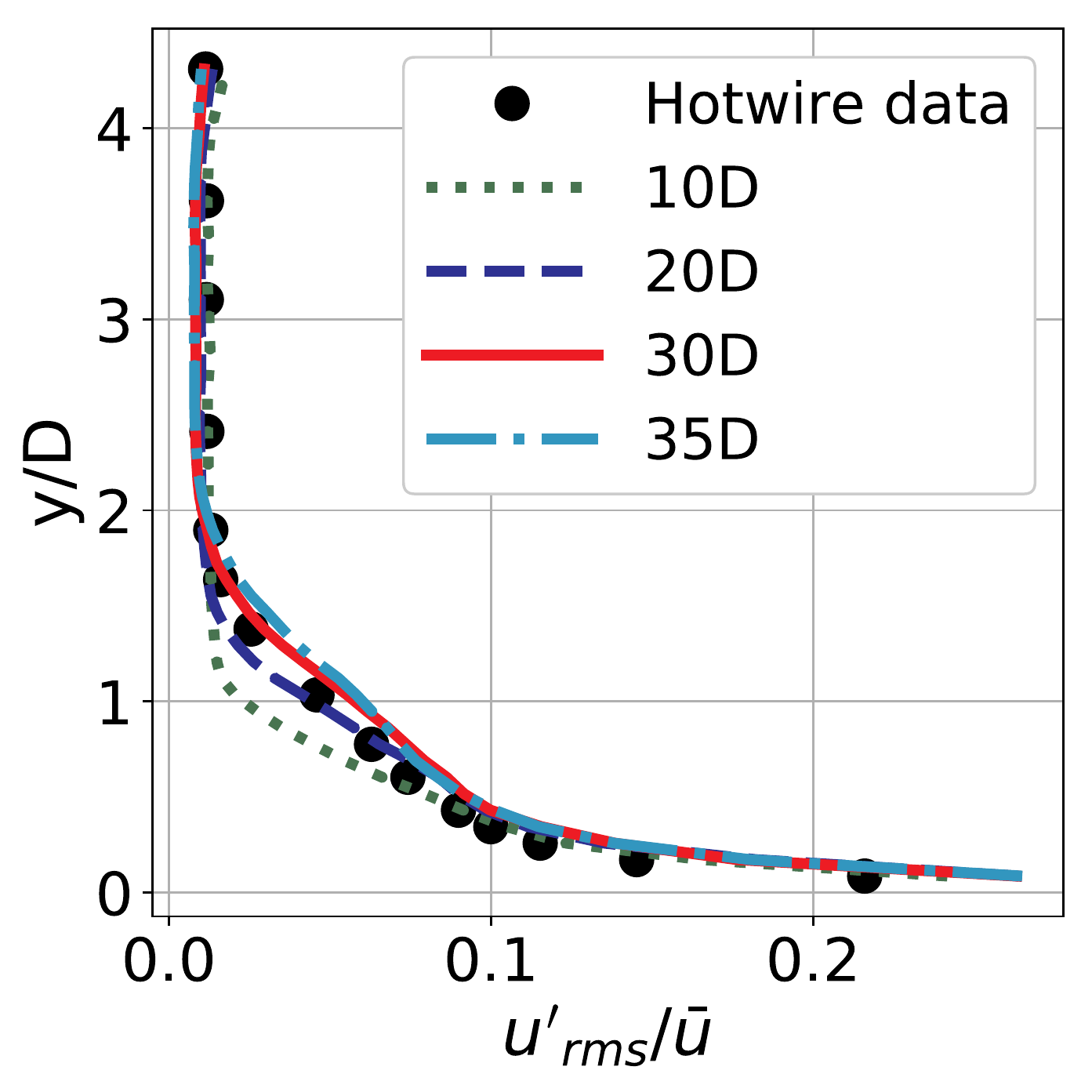}}}%
    \subfloat[]{{\includegraphics[width=55mm]{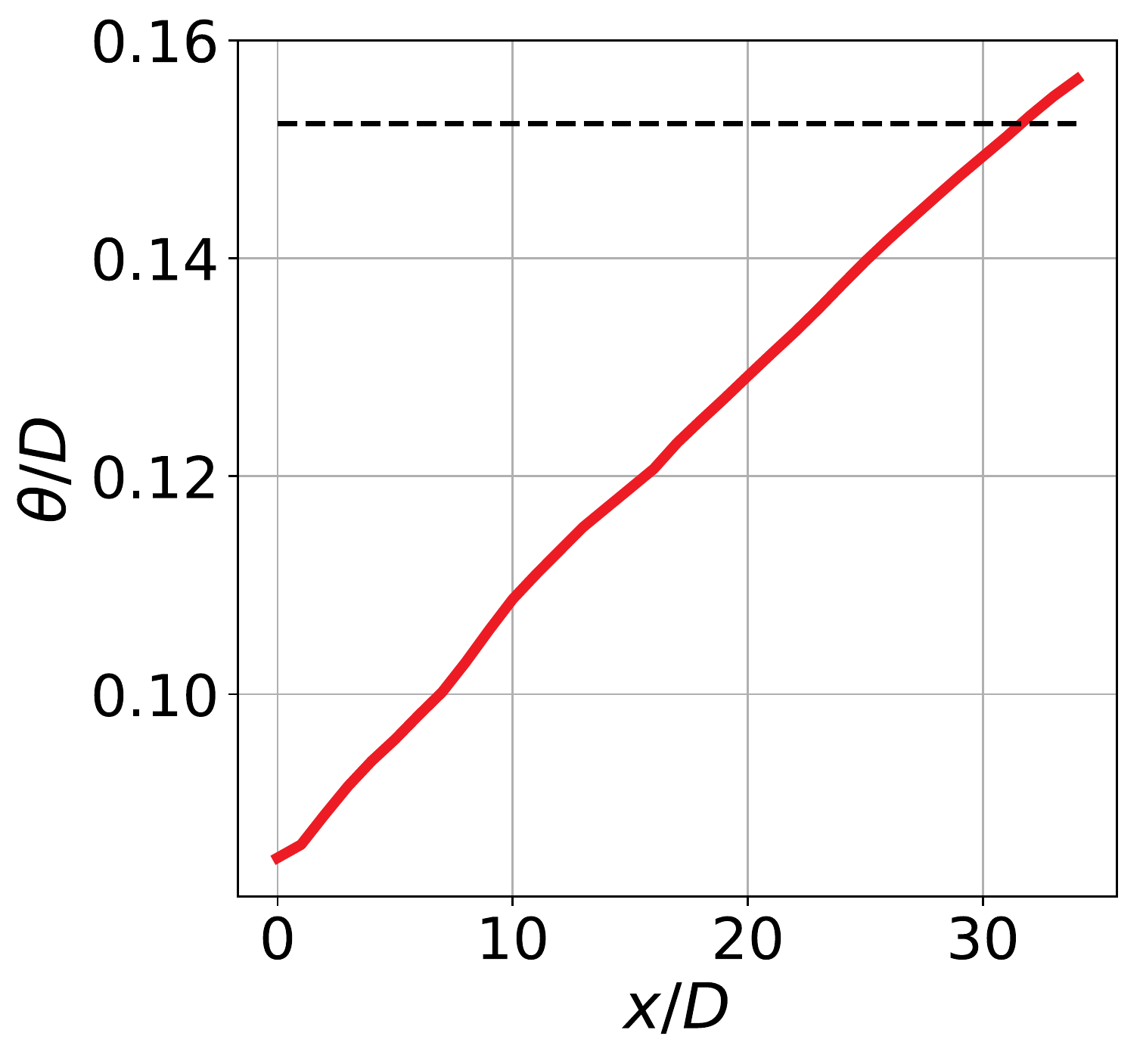}}}%
  \end{center}
  
\caption{Comparison between preliminary channel flow LES and the hot-wire experimental data. (\textit{a}) and (\textit{b}) show vertical profiles of mean streamwise velocity and streamwise velocity rms respectively, from the hot-wire and from different stations denoted by their distance from the channel inlet. (\textit{c}) plots momentum thickness versus streamwise distance for the channel flow LES, together with the value measured with the hot-wire at the hole location (dashed line); note that the channel flow LES achieves the hot-wire value of $\theta / D$ at roughly $30D$ downstream of the inlet.}
\label{fig-9-inlet}
\end{figure}

The boundary conditions at all walls are no-slip for the velocity field and zero-flux (adiabatic) for the scalar concentration, which match the physical behavior of the experiment. At the bottom wall and the jet wall, the equations are solved directly without wall functions since the mesh resolution is fine enough. At all other solid boundaries, the wall functions of \citet{cabot2000approximate} are employed for the velocity field.

Time-varying inlet conditions for the velocity at the channel and plenum feed inlet are generated using a method similar to \citet{xie2008efficient}. The inlet generator requires the mean velocity and Reynolds stress 2D profiles, together with a length scale and a time scale. The goal of the present simulations is to match the experimental setup; thus, these quantities must be carefully selected to recover experimental profiles.

To specify inlet conditions for the full simulation, an iterative procedure was employed which involved running preliminary Large Eddy Simulations of a developing square channel flow until a good inlet condition was achieved. Importantly, the channel flow LES uses the same mesh configuration as used in the equivalent region of the final jet in crossflow LES. Hot-wire measurements of mean streamwise velocity, $\bar{u}$, and root-mean-square of the fluctuating streamwise velocity, $u'_{rms}$ are used for validation. The method is described in Tab.~\ref{tab-2-inletalgorithm}:

The final inlet condition parameters, ultimately used for all three jet in crossflow LES's, were obtained after several such iterations. They consist of 2D mean profiles based on the 1D hot-wire data, re-scaled to a thinner boundary layer (by a factor of 0.55). The $u'_{rms}$ profile was magnified by a factor of $1.5$ in the boundary layer and by a factor of $5$ in the freestream, because the synthetic turbulence tends to decay rapidly in the first few hole diameters after the inlet. The turbulent length scale required by the synthetic turbulence generator was set to $0.7D$. Fig.~\ref{fig-9-inlet}(a)-(b) shows the hot-wire data along with profiles of $\bar{u}(y)$ and $u'_{rms}(y)$ from different stations downstream of the channel inlet. The agreement for both mean velocity and rms velocity profiles is best for an inlet plane $30D$ upstream of the hole position. Furthermore, the momentum thickness $30D$ downstream of the inlet plane is $0.149D$, very close to the value of $0.15D$ measured in the experiment, as shown in Fig.~\ref{fig-9-inlet}(c). Therefore, we choose to set the inlet plane $30D$ upstream of the hole location.

Other preliminary tests showed that changing the velocity profile at the plenum inlet had small effect on the jet-crossflow interaction. Therefore, a uniform mean velocity and a uniform, low level of isotropic turbulence fluctuations were used in this inlet.

\section{Results and validation}

\subsection{Mean scalar field}

\begin{figure}
  \begin{center}
  \includegraphics[width = 85mm] {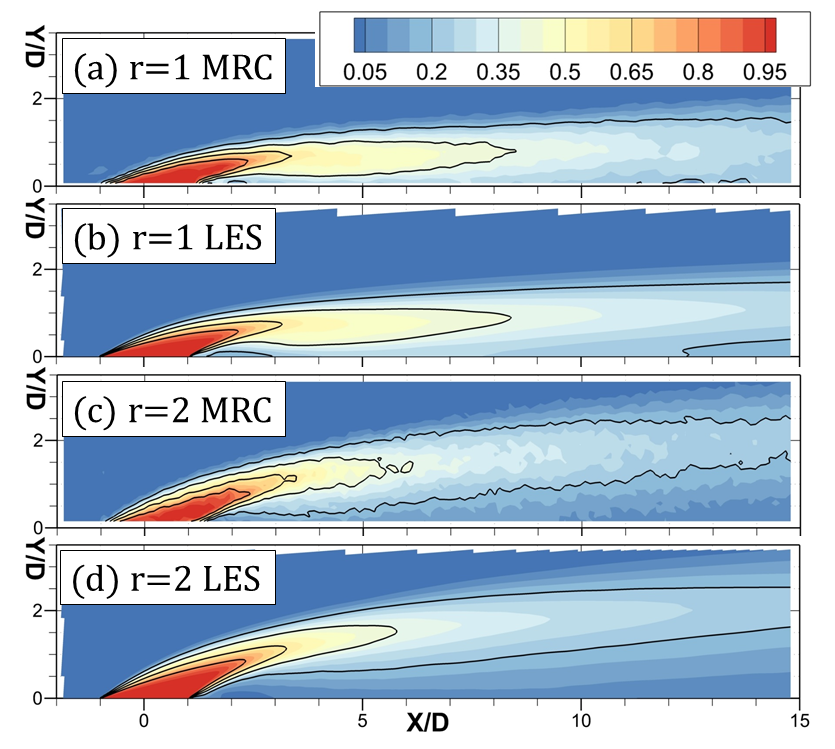}
  \end{center}
\caption{Mean scalar concentration $\bar{c}$ at the $z=0$ spanwise plane (symmetry plane for channel and jet). Results are shown for two velocity ratios, $r=1$ and $r=2$, and for both experiments (MRC) and simulations (LES). Lines indicate isocontours of $\bar{c}=0.2, 0.4, 0.6, 0.8$.}
\label{fig-10-concz0}
\end{figure}

Fig. \ref{fig-10-concz0} shows mean concentration results in the symmetry plane ($z=0$) for two velocity ratios, $r=1$ and $r=2$. In this view, the jet meets the main channel at $y=0$, between $x/D=-1$ and $x/D=1$. Qualitatively, one notes the different behavior of the two configurations. When $r=1$, the scalar concentration initially separates from the bottom wall, but re-attaches at about $x/D=3$ and stays attached all the way to the outlet. At $r=2$, the vertical momentum of the jet is sufficiently high for it to stay completely detached throughout the whole domain. It is not shown here, but the LES results from the $r=1.5$ jet show that it is also completely detached; this implies that for the present geometry, the jet becomes fully detached starting at a critical velocity ratio somewhere between $r=1$ and $r=1.5$.

This regime, in which the jet detaches from the wall however briefly, is the one in which turbulence models fail to predict adequate mean concentrations in the context of RANS simulations \citep{bogard_review}. Comparing the experimental data to the present simulations (Figs.~\ref{fig-10-concz0}(a) to (b) and (c) to (d)) yields excellent agreement in the symmetry plane, which shows that LES is a tool much better suited at capturing the scalar mixing than RANS models. 

\begin{figure}
  \begin{center}
  \includegraphics[width = 60mm] {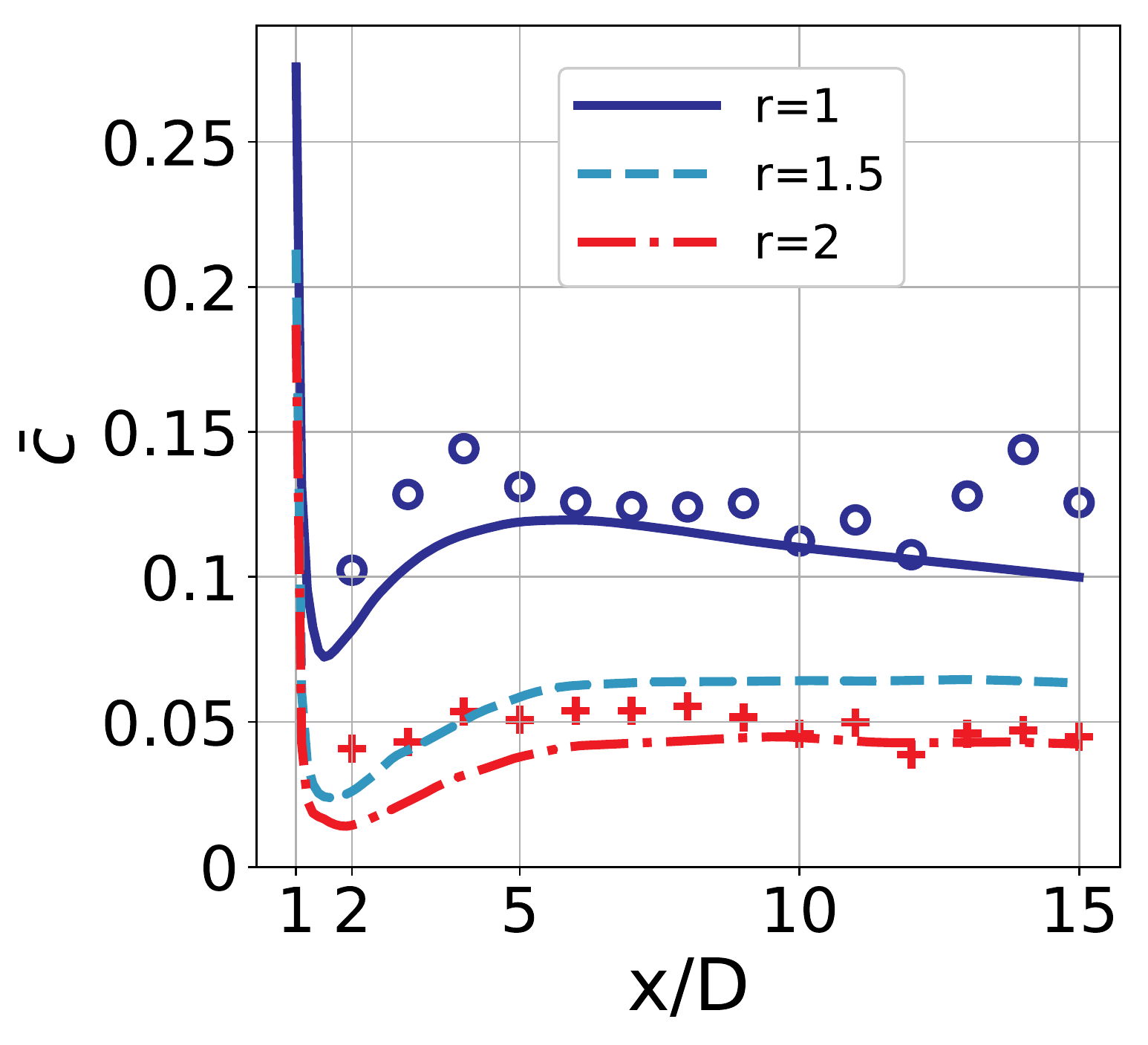}
  \end{center}
\caption{Mean scalar concentration at the bottom wall, averaged between $z/D=-1$ and $z/D=1$. Solid lines are from the LES and symbols are from MRC measurements.}
\label{fig-11-concy0}
\end{figure}

The mean scalar concentration evaluated at the wall is of particular interest in film cooling applications, where it is referred to as adiabatic effectiveness. The solid lines in Fig.~\ref{fig-11-concy0} show the mean scalar concentration at the wall in the LES, averaged between $z/D=-1$ and $z/D=1$, for each velocity ratio. The main feature of jets that separate from the wall right after injection is the sharp drop in the averaged adiabatic effectiveness after injection, as can be seen for all three velocity ratios. RANS simulations employing typical turbulent mixing models fail to predict this behavior (\citet{milani2018approach}). Since the $r=1$ jet re-attaches to the wall, its averaged adiabatic effectiveness recovers to much higher levels, and does so over a short streamwise distance. That does not happen with the other two cases: their adiabatic effectiveness curves are similar, and only increase mildly after the initial drop-off. The experimental results are also shown as symbols for $r=1$ and $r=2$. Since the MRI techniques are not appropriate for wall measurements, the experimental values used in Fig.~\ref{fig-11-concy0} come from the first measurement point in the fluid above the wall. This biases the results to be slightly higher than the actual adiabatic effectiveness and is prone to higher overall noise levels due to partial volume effects. With those caveats, the agreement between LES and MRC in Fig.~\ref{fig-11-concy0} is good. This comparison illustrates the value of enriching: the experiments, though comprehensive, cannot directly produce the adiabatic effectiveness, so a detailed and validated simulation is used to fill in the gap for the exact geometry.

\begin{figure}
  \begin{center}
  \includegraphics[width = 160mm] {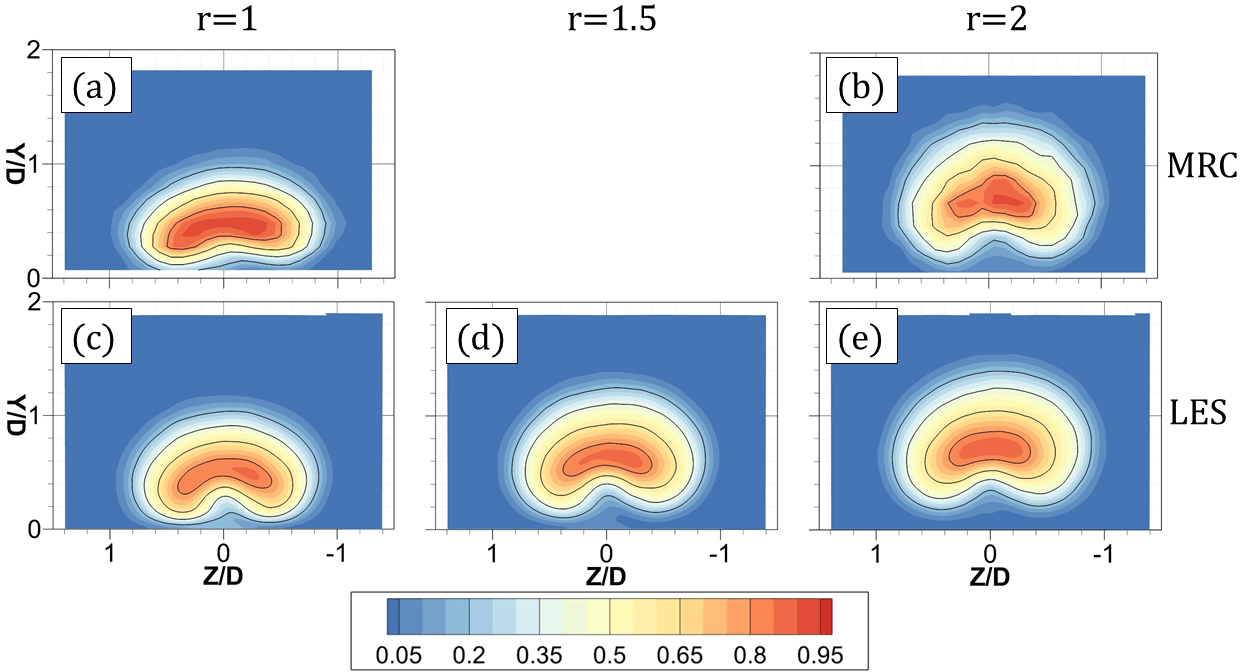}
  \end{center}
\caption{Mean scalar concentration values at a streamwise plane located at $x/D=2$ for different velocity ratios. The first row, (\textit{a})-(\textit{b}), contains MRC results, and the second row, (\textit{c})-(\textit{e}), contains LES results. Lines indicate isocontours of $\bar{c}=0.2, 0.4, 0.6, 0.8$.}
\label{fig-12-concx2}
\end{figure}

\begin{figure}
  \begin{center}
  \includegraphics[width = 160mm] {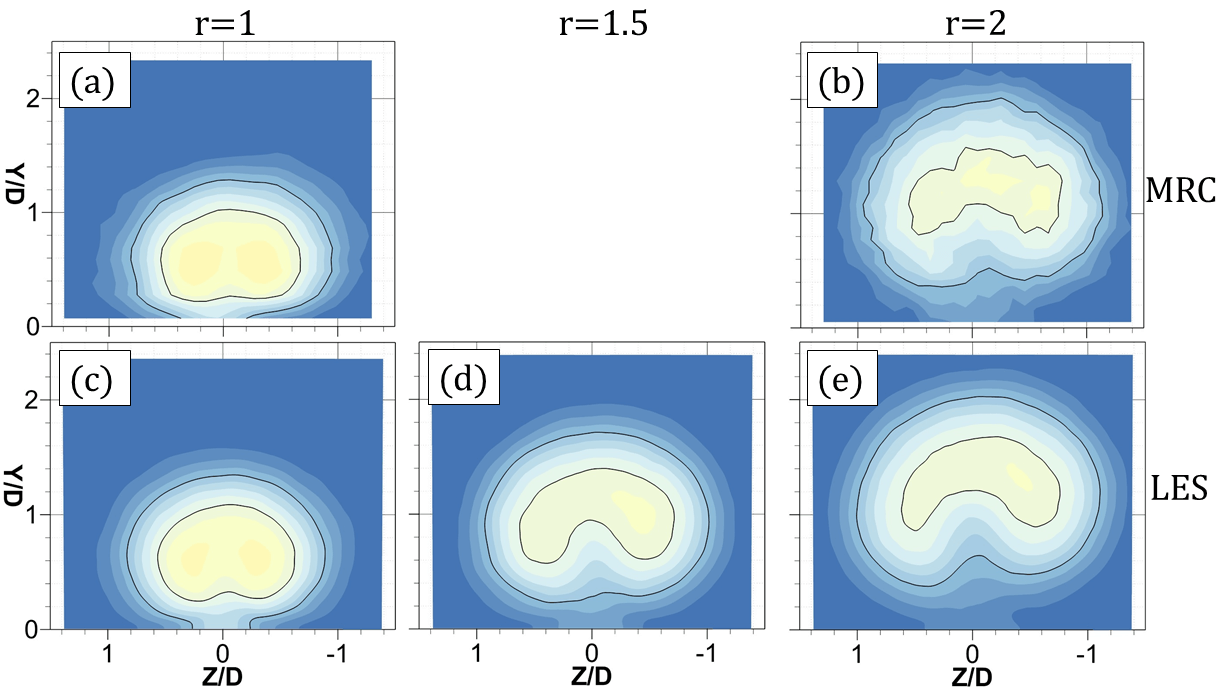}
  \end{center}
\caption{Mean scalar concentration values at a streamwise plane located at $x/D=5$ for different velocity ratios. The color scheme and isocontours are the same as in Fig.~\ref{fig-12-concx2}.}
\label{fig-13-concx5}
\end{figure}

Figs.~\ref{fig-12-concx2} and \ref{fig-13-concx5} show mean concentration results in streamwise (y-z) planes, in which the main flow direction is out-of-the-page. They are located, respectively, at $x/D=2$ and $x/D=5$. The dominating feature of such mean concentration fields is their kidney shape, previously documented in the literature (e.g. \citet{smith1998mixing}). This characteristic shape is caused by the deformation of the jet due to the counter-rotating vortex pair. As expected, higher velocity ratio causes the concentration profiles to be located farther from the bottom wall. Another effect of the velocity ratio regards the jet structure. Increasing $r$ causes the jet area to increase, because the fluid released by the hole must decelerate more as it exchanges momentum with the freestream, and thus must occupy a larger cross-sectional area by conservation of mass. The contours show that most of that increase in area is due to increase in the vertical extent of the jet cross-section: while the jet width increases only mildly from $r=1$ to $r=2$, the jet height increases significantly. This is particularly evident in Fig.~\ref{fig-13-concx5}. One possible explanation is that, as the jet trajectory moves further away from the wall, there is more space between the jet core and the bottom wall. The turbulent transport in the vertical direction increases since the eddies are farther away from the wall. The spanwise transport, on the other hand, is not magnified by any of these factors.

Finally, Figs.~\ref{fig-12-concx2} and \ref{fig-13-concx5} suggest that there is mild asymmetry in the jet structure about the $z=0$ plane, as has been discussed by authors such as \citet{smith1998mixing}. In the present setup, it would be chiefly caused by asymmetry in the plenum feed (see Fig.~\ref{fig-4-mesh}), not manufacturing tolerances or the main channel inlet.

\begin{figure}

  \begin{center}
    \subfloat[]{{\includegraphics[width=75mm]{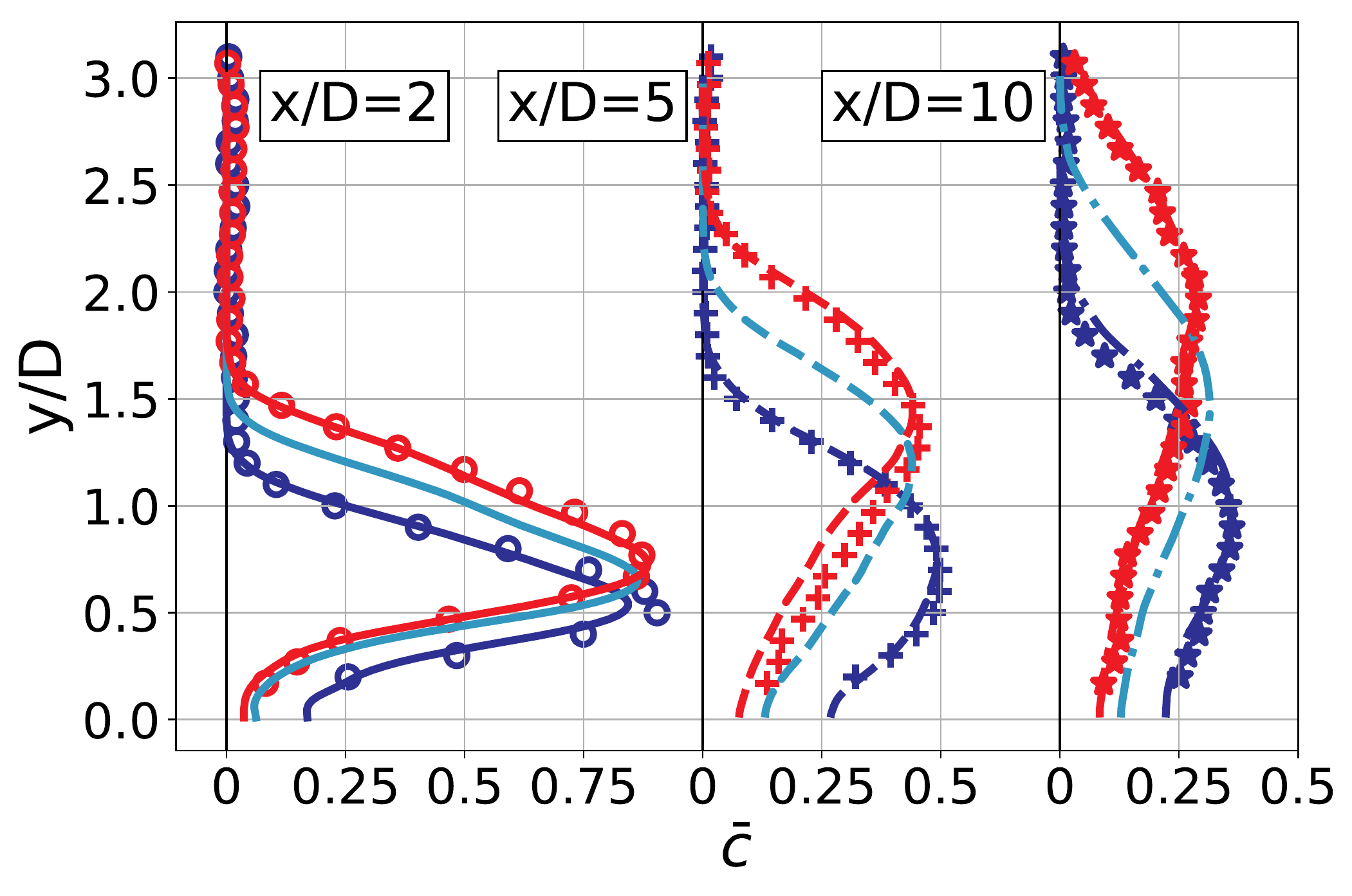}}}%
    \subfloat[]{{\includegraphics[width=66mm]{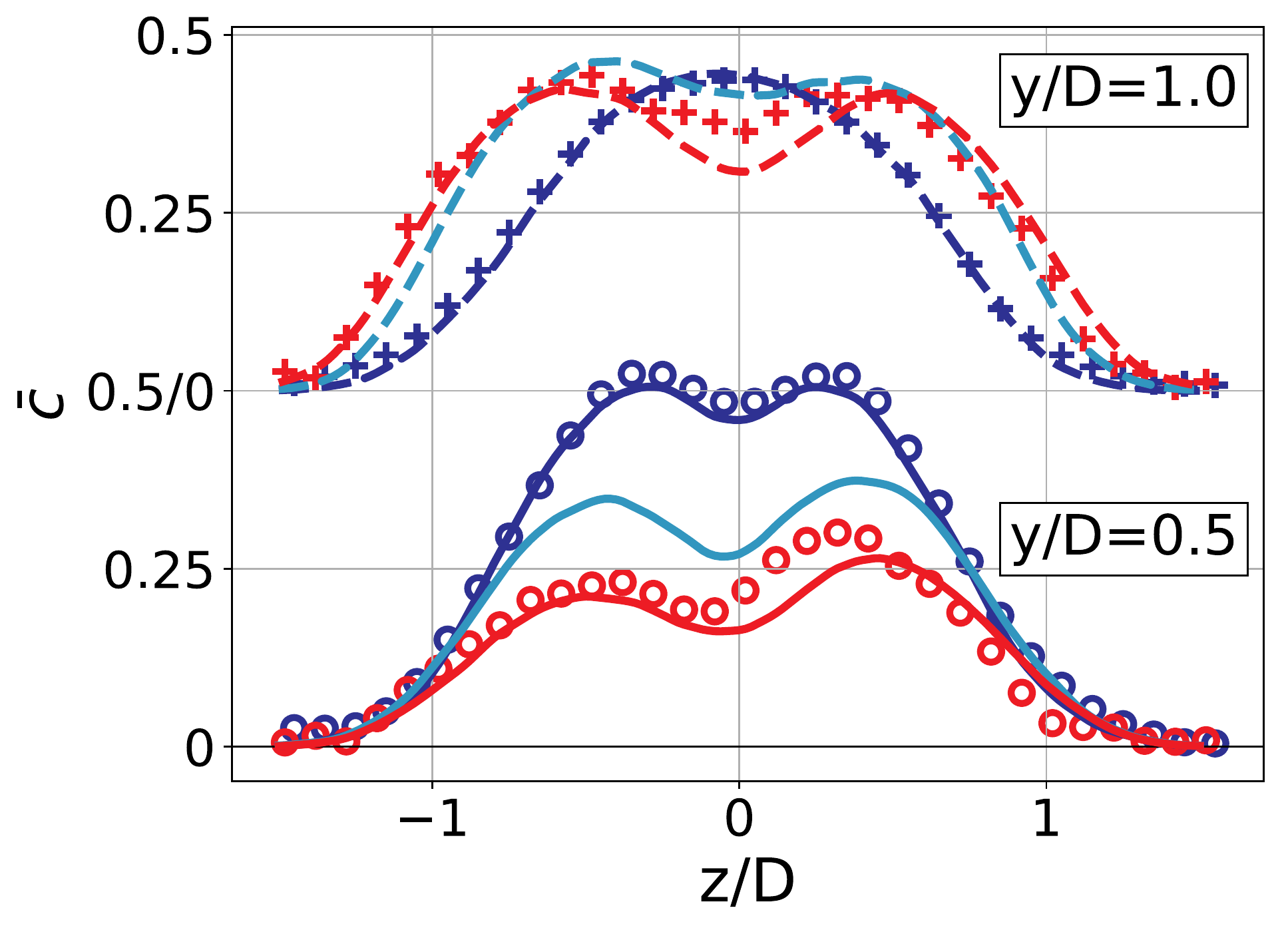}}}%
  \end{center}
  
\caption{1-D profiles of mean concentration extracted from LES (lines) and MRC (symbols). All velocity ratios are shown: $r=1$ in blue, $r=1.5$ in cyan, and $r=2$ in red. (\textit{a}) shows vertical profiles at the centerline ($z=0$) and at different streamwise positions; (\textit{b}) contains horizontal profiles at $x/D=5$ at different heights.}
\label{fig-14-profilesc}
\end{figure}

For a more quantitative comparison between simulation and experiments and between different velocity ratios, Fig.~\ref{fig-14-profilesc} presents vertical and horizontal profiles of mean concentration. Fig.~\ref{fig-14-profilesc}(a) shows vertical profiles at the centerline at different streamwise positions. It shows that at $x/D=2$ the jet core is still cohesive, and increasing the value of $r$ seems to move the profile up without changing its shape. Further downstream, turbulent mixing acts to spread the jet, and its effects are more pronounced as $r$ increases. Fig.~\ref{fig-14-profilesc}(b) shows double-peaked profiles due to the kidney shape in Fig.~\ref{fig-13-concx5}, and confirms that the width of the jet, measured by its spanwise profile, does not change much across velocity ratios.

The comparison between experiments and simulation in Figs.~\ref{fig-12-concx2}, \ref{fig-13-concx5}, and \ref{fig-14-profilesc} shows that, in general, the mean concentration predictions agree well with the data. Some discrepancy is observed between Figs.~\ref{fig-12-concx2}(a) and \ref{fig-12-concx2}(c), where the LES seems to slightly underpredict the maximum concentration and the width of the scalar profile. These particular plots should be the most sensitive to small differences in operating conditions and inlet conditions between simulation and experiment, which probably account for the discrepancies observed. For $r=2$ in Fig.~\ref{fig-14-profilesc}(b), the LES and MRC qualitatively agree on the asymmetry across the double peak (showing higher $\bar{c}$ on the $z>0$ side at the lower profile) despite their slight quantitative mismatch, which suggests that simulation has enough fidelity to capture the effect of the plenum feed on the resulting concentration field in the main channel.

\subsection{Mean velocity field}

\begin{figure}
  \begin{center}
  \includegraphics[width = 100mm] {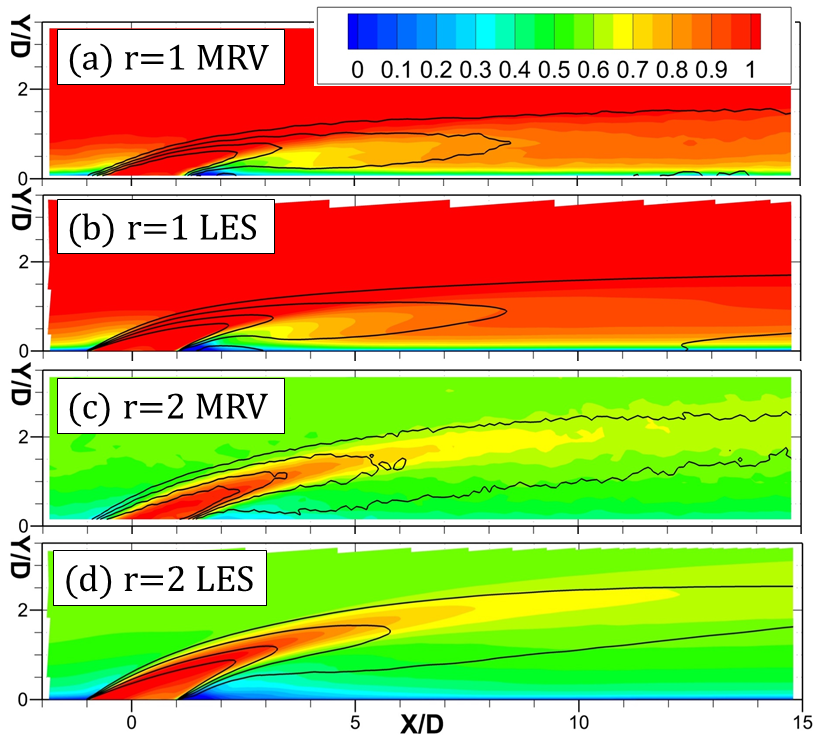}
  \end{center}
\caption{Mean streamwise velocity normalized by jet bulk velocity, $\bar{u}/(rU_c)$, at the symmetry plane. Results are shown for $r=1$ and $r=2$, and from data (MRV) and simulations (LES). Lines indicate isocontours of $\bar{c}=0.2, 0.4, 0.6, 0.8$, same as in figure~\ref{fig-10-concz0}}
\label{fig-15-uz0}
\end{figure}

In this subsection, the mean velocity data from the MRV and LES are presented and discussed. For adequate comparison of the interaction between jet and crossflow across different velocity ratios, it is most convenient to normalize the velocity results by the jet bulk velocity, $rU_c$. Fig.~\ref{fig-15-uz0} shows mean streamwise velocity in $z=0$ planes, superimposed on isocontour lines of mean concentration. In all cases, one notes that before meeting the jet, the incoming boundary layer on the channel wall has thickness comparable to the hole diameter. It grows immediately upstream of the jet due to the adverse pressure gradient induced by the jet's blockage. \citet{fric1994vortical} argued that this process is important for the resulting downstream vortical structure. An observation from the $r=2$ data is that the point of highest scalar concentration is closer to the wall then the point of highest mean streamwise velocity, which appears as an apparent mismatch between the lines and the color contours. This shows that the jet, as measured by the velocity field, penetrates deeper into the crossflow than the jet measured by the concentration field. This effect is also observed using different metrics by \citet{coletti2013turbulent}.

\begin{figure}
  \begin{center}
  \includegraphics[width = 160mm] {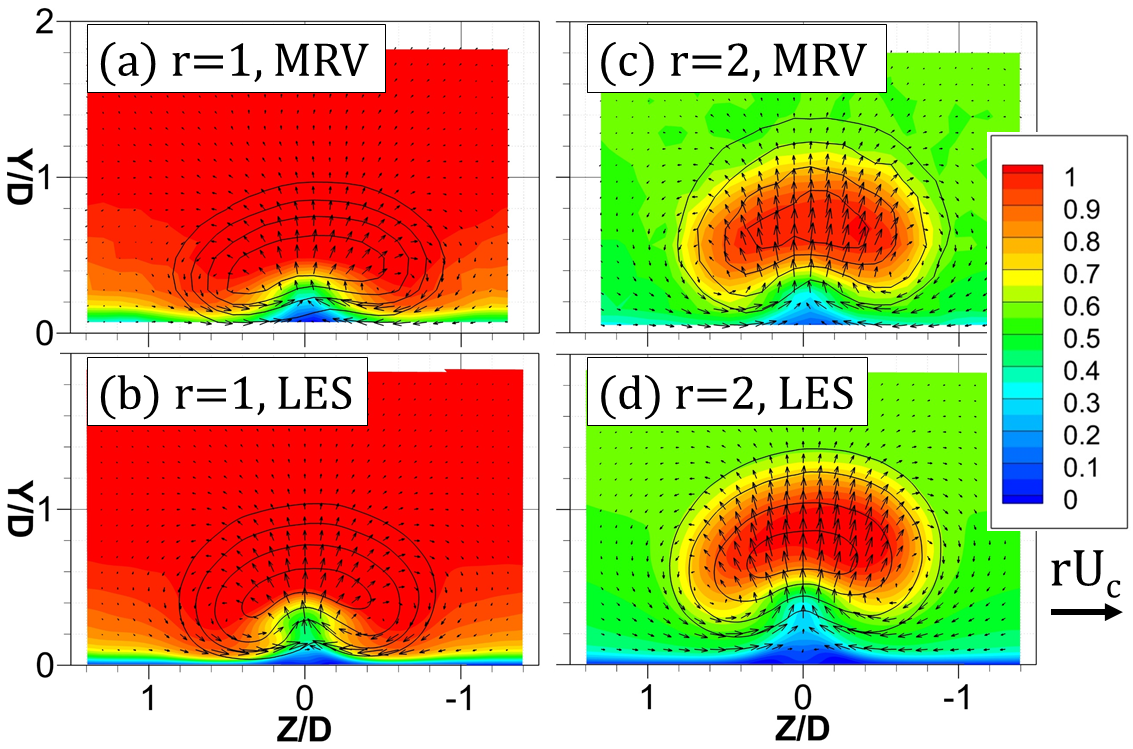}
  \end{center}
\caption{Streamwise planes located at $x/D=2$ showing mean velocities normalized by $r$. Color contours show streamwise velocity levels, $\bar{u}/(rU_c)$, and vectors show in-plane velocities (a reference vector is provided on the bottom right). Black lines denote isocontours of mean concentration, identical to Fig.~\ref{fig-12-concx2}.}
\label{fig-16-velx2}
\end{figure}

Mean secondary flows are significant in jets in crossflow. Fig.~\ref{fig-16-velx2} shows streamwise planes at $x/D=2$ with information on mean streamwise velocity, mean secondary flows, and mean concentration. The counter rotating vortex pair (CVP) is the most conspicuous feature of this flow, clearly shown by the vectors of in-plane velocity. Instantaneously, the CVP is present, but highly unsteady and asymmetric as discussed by \citet{smith1998mixing}; unlike other vortical structures, it also appears in the mean flow. The CVP direction is common-up, and right after injection the center of the vortices is located at approximately $z/D = \pm 0.5$, i.e. above the side edges of the holes. The superposition of the in-plane velocity with the mean scalar concentration shows clearly that the CVP acts to distort the scalar field, creating the kidney shape shown in Fig.~\ref{fig-12-concx2}. It also acts to sweep in mainstream flow under the jet (which is responsible for the dip in wall concentration shown in Fig. ~\ref{fig-11-concy0}), and push up low velocity fluid located in the boundary layer and under the jet (a phenomenon which is visible in Figs.~\ref{fig-15-uz0} and \ref{fig-16-velx2}).

\begin{figure}

  \begin{center}
    \subfloat[]{{\includegraphics[width=97mm]{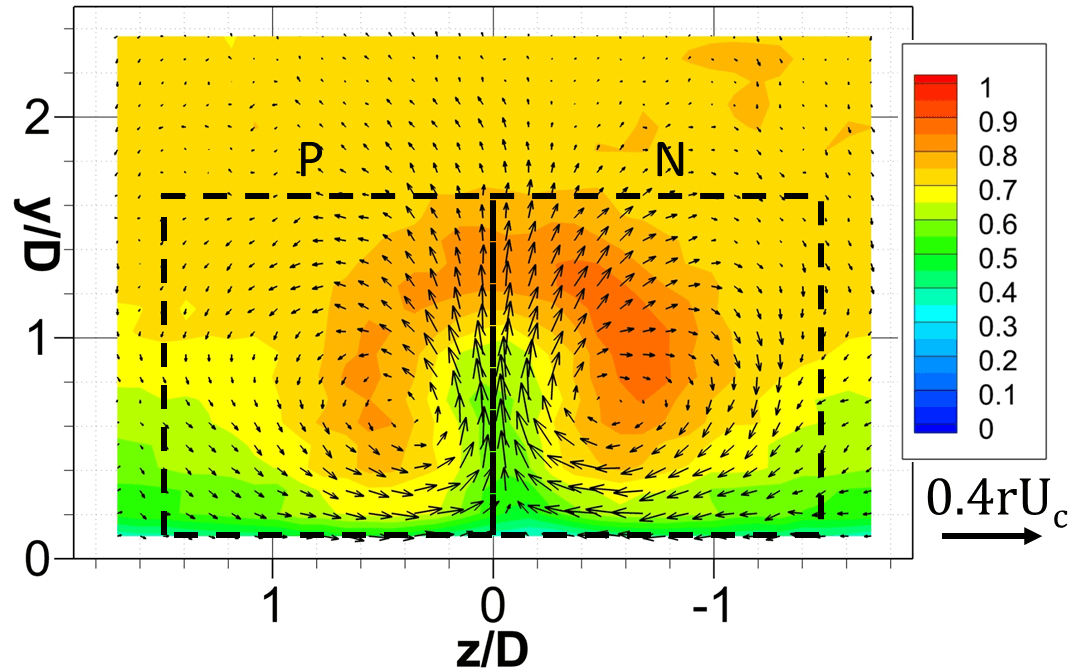}}}%
    \subfloat[]{{\includegraphics[width=65mm]{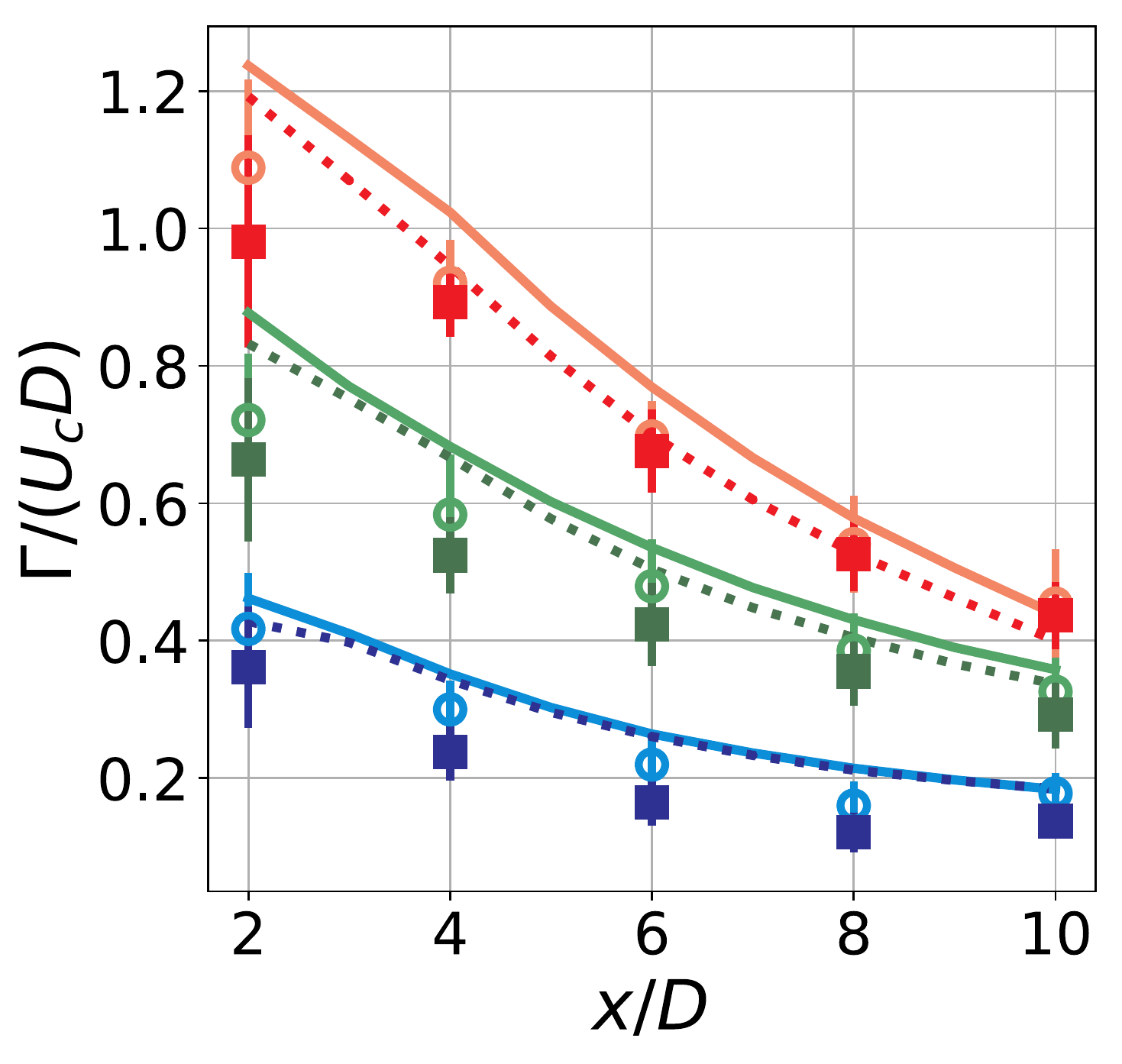}}}%
  \end{center}
  
\caption{(a) Streamwise plane located at $x/D=5$ showing mean velocities normalized by $r$ in the $r=1.5$ case, with two loops (positive $P$ and negative $N$) showing where the circulation is calculated. (b) Dimensionless circulation as a function of streamwise location in each velocity ratio ($r=1$ in blue, $r=1.5$ in green, and $r=2$ in red). LES data are shown in lines and MRI data are shown in symbols. Solid lines and circles denote negative side $N$, and dashed lines and squares denote positive side $P$.}
\label{fig-17-circulation}
\end{figure}

To further understand and validate the simulation results, it is useful to look at quantitative scalar metrics derived from the 3D data. In the context of mean secondary flows, an important metric is the circulation $\Gamma$, which consists of the line integral of the mean velocity field along a right-handed closed path. At different axial planes, we can define closed lines that form a square and encompass each vortex of the CVP as shown in Fig.~\ref{fig-17-circulation}(a) and calculate the circulation around them. The square has side $1.5D$ and has a corner at $z/D=0$ and $y/D=0.15$; the latter was chosen because MRV data are unreliable closer to the wall, and we are interested in the total circulation from the vortices (not the boundary layer). Since the vortex is common-up, the circulation will be positive in the side marked by $P$ and negative in the side marked by $N$; if the flow were perfectly symmetric, it would have the same magnitude on both sides.

Fig.~\ref{fig-17-circulation}(b) shows dimensionless circulation $\Gamma / (U_c D)$ calculated over the curve shown in Fig.~\ref{fig-17-circulation}(a) defined at different streamwise locations. The lines indicate LES data and the symbols indicate MRV data. To obtain error bars on the MRV calculation, the uncertainty due to possible misalignment of the data was taken into account. The actual curve over which the circulation is to be calculated was shifted randomly in all three directions by up to a single MRI voxel ($0.1D$) fifty times and the error bars indicate two standard deviations around the mean result. The plot contains the absolute value of the circulation in each side marked in Fig.~\ref{fig-17-circulation}(a) ($N$ and $P$). As we can see from the results, the circulation is almost symmetric, with a clear tendency of higher values on the $N$ side (possibly due to asymmetry of the plenum feed). It also seems to increase superlinearly with the velocity ratio $r$, since values at $r=2$ are more than double those of $r=1$. Regarding the agreement between simulation and experiment, the LES reproduces well the trends of the experiment, including the direction of asymmetry, the streamwise decay, and the velocity ratio dependency. However, the simulations seems to consistently overpredict $\Gamma$, sometimes beyond what is explained by misalignment errors alone. This could be attributed to other experimental uncertainties (such as velocity ratio) or to possible differences in the in-hole flow between experiment and simulation, which will be discussed next.

\begin{figure}
  \begin{center}
  \includegraphics[width = 150mm] {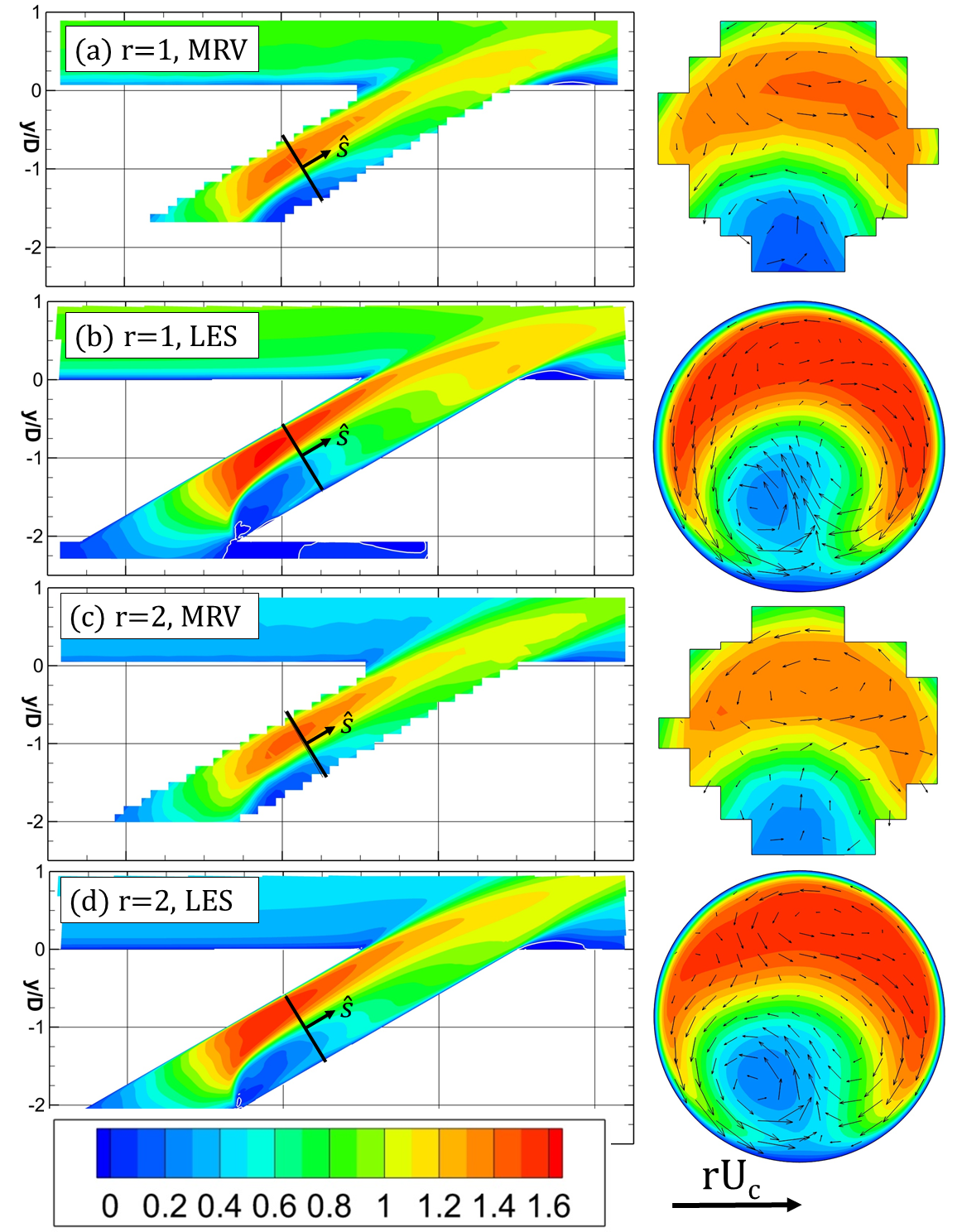}
  \end{center}
\caption{Plots showing in-hole velocity for (\textit{a}-\textit{b}) $r=1$ and (\textit{c}-\textit{d}) $r=2$. The contours are of mean axial velocity along the hole $\bar{u}_s/(rU_c)$. On the left, the jet symmetry plane is shown. On the right, axial planes are shown in the middle of the hole. Note that the axial planes also contain vectors showing mean in-plane velocity, and a reference vector is provided next to the contour legend.}
\label{fig-18-velhole}
\end{figure}

An important feature of the present flow is the flow inside the hole, before the jet meets the crossflow. In idealized conditions, fully developed pipe flow would prevail at the jet exit, which is the case in the experiment of \citet{su2004simultaneous} and the simulation of \citet{muppidi2008direct}. However, in practical applications, the jet fluid may be expelled by a short pipe, or the pipe could be fed in such a way as to induce important irregularities in the flow. This is often the case in film cooling, where the coolant jet comes from a plenum at the base of a short hole. In order to study possible effects from that, the present work deals with a short hole, of length $4.1D$, fed from a more realistic plenum. 

Fig.~\ref{fig-18-velhole} shows experimental and numerical results for the mean velocity within the hole in the $r=1$ and $r=2$ cases. The coordinate system used is aligned with the inclined hole, where $\hat{s}$ is the unit vector pointing the the axial direction. The contour colors, in all plots, denote mean fluid velocity in the $\hat{s}$ direction, or $\bar{u}_s$. The axial plane is located in the midpoint along the hole between the plenum and the channel bottom wall, and its location is marked as a black line in the centerplane. Vectors in the axial plane show mean in-plane velocity. 

It is clear from these plots that the fluid coming from the plenum must turn along the right corner of the hole, causing flow separation on that side. That phenomenon is responsible for the highly non-uniform axial velocity observed in the axial planes shown: much higher velocity on the top, and lower velocity in the wake of the separation bubble. If the pipe were long enough, the flow would develop towards Poiseuille flow; since the pipe is only $4.1 D$ in length, this non-uniform flow persists all the way to injection. The axial planes also show significant secondary flows, with magnitude of up to about 20\% of the bulk velocity in the hole. In both cases, there is a counter-clockwise vortex on the top of the planes, and flow along the walls from the top to the bottom. 

At first, it might seem intuitive to think that changing velocity ratio from $r=1$ to $r=2$ would impact the flow structure in the hole. That is because the Reynolds number based on hole bulk velocity increases by a factor of two, and the blockage effect as seen by the hole caused by the jet meeting the crossflow becomes less significant as $r$ increases. However, both of these effects seem to be small because Fig.~\ref{fig-18-velhole} shows that the velocity field non-dimensionalized by $rU_c$ is mostly insensitive to $r$ in this range. The only discernible difference is that the wake of the separation bubble seems to recover slightly faster for $r=1$ compared to $r=2$, an effect captured both in the LES and the MRV.

It is also important to comment on the agreement between the LES and the MRV within the hole. Qualitatively, the plots agree well, including the trends of the secondary flows and the location and size of the separation and its wake. Quantitatively, however, the agreement is not as good as that shown in other regions of the flow.
MRV fails to provide high quality data close to solid walls because of signal loss due to partial volume effects. Signal loss is also observed in regions of excessive turbulence. These effects combined with the low experimental resolution at the hole, which is not enough to capture the sharp velocity gradients shown in the LES, greatly increase uncertainty there. Thus, for the present MRI setup, analysis of the data should be limited to broad qualitative features in the hole. Given that the LES matches the experiments elsewhere, this is a region where the LES results are more trustworthy than the MRI data and can be used to enrich them.

\subsection{Turbulence data}

Two examples of enrichment of MRI data were shown in the mean data, namely concentration at the wall and velocity in the hole. Another broad category of numerical information that complements the MRI data is turbulence statistics, which can be calculated due to the unsteady nature of the LES but are not available in MRI data. The present subsection briefly exemplifies these results.

\begin{figure}
  \begin{center}
  \includegraphics[width = 160mm] {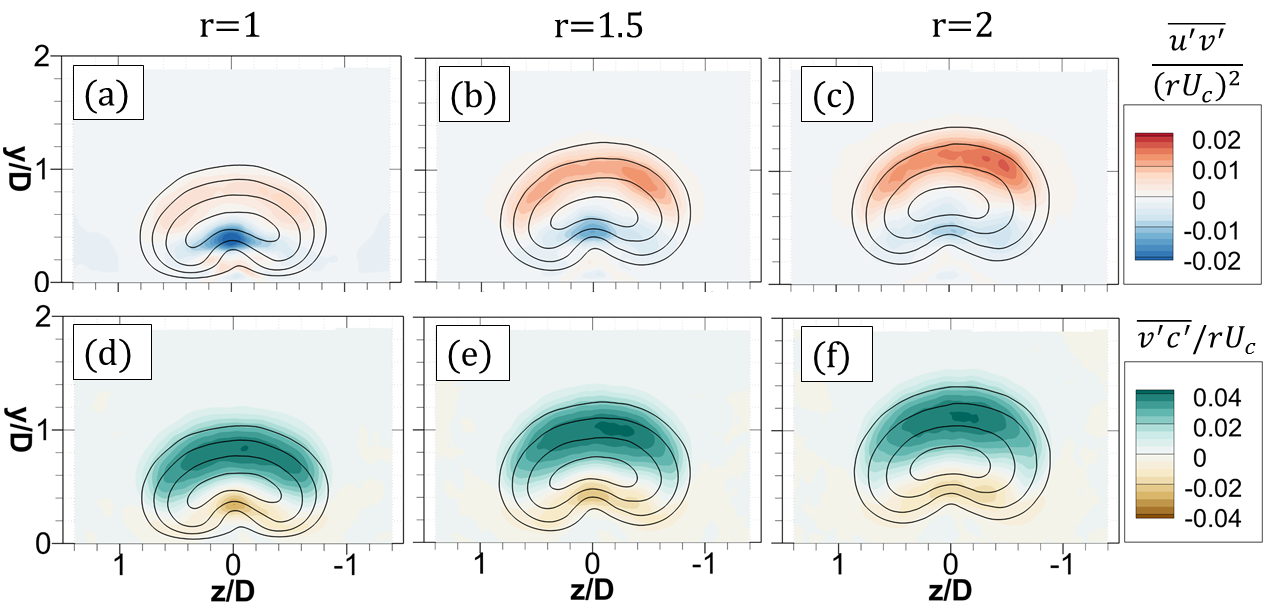}
  \end{center}
\caption{Contours of $\overline{u'v'}$ (top row) and $\overline{v'c'}$ (bottom row) provided by the LES non-dimensionalized by the jet velocity, $U_j=rU_c$. The plots show streamwise ($y-z$) planes located at $x/D=2$ and black lines denote isocontours of mean concentration at $\bar{c}=0.2, 0.4, 0.6, 0.8$.}
\label{fig-19-x2turbulence}
\end{figure}

Fig.~\ref{fig-19-x2turbulence} shows two turbulence correlations that play an important role in the Reynolds-averaged equations for velocity and concentration. It contains $y-z$ planes at $x/D=2$ for all three velocity ratios. Figs.~\ref{fig-19-x2turbulence}(a)-(c) show one component of the Reynolds stress, $\overline{u'v'}$, with both velocity factors non-dimensionalized by the jet velocity. Figs.~\ref{fig-19-x2turbulence}(d)-(f) show the vertical turbulent scalar flux, $\overline{v'c'}$, with the velocity scaled by the jet velocity. The non-dimensionalization was chosen in an attempt to represent all velocity ratios in the same contour. 

Broadly, the two turbulence quantities behave similarly: they are positive on the top shear layer, negative on the bottom half of the jet, and close to zero at the jet core. However, they seem to scale differently as the velocity ratio changes: $\overline{u'v'}$ changes more drastically (with the positive regions intensifying and the negative regions lessening), while the values of $\overline{v'c'}$ (scaled by $r$) are similar across the three simulations. This might be due to changing mean profiles: as can be seen in Fig.~\ref{fig-15-uz0}, the structure of mean concentration gradients does not change much from $r=1$ to $r=2$, but the mean velocity gradients experience quantitative change because in the former the jet core is slower than the freestream and in the latter it is faster. Note also the counter-intuitive behavior of $\overline{u'v'}$ and $\overline{v'c'}$ under the jet: this is a location where $\partial \bar{u} / \partial y > 0$ and $\partial \bar{c} / \partial y > 0$, but $\overline{u'v'}$ and $\overline{v'c'}$ are weakly positive. This suggests that linear eddy viscosity models would struggle to model the turbulence in this region.

\section{Conclusion}

In the present work, 3D MRI experiments and highly resolved Large Eddy Simulations were performed on the same geometry, an inclined jet in crossflow, with three distinct velocity ratios ($r=1$, $r=1.5$, and $r=2$). The flow is incompressible, the Reynolds number based on the jet velocity varies between $Re_D = 2,900$ and $Re_D = 5,800$, and the concentration of a passive scalar is also tracked. An important contribution of this work is to document the effort to carefully match operating conditions between MRI experiments and high fidelity simulations, especially that of setting consistent inlet conditions. When this is done properly, good agreement can be expected across the whole 3D domain; then, since the datasets credibly represent the same flow, we say that the LES enriches the MRI dataset with information that the experiment is unable to provide. Future workers who perform MRI experiments can follow some of the present guidelines to complement their experimental data with Large Eddy Simulations, allowing them access to turbulent statistics and near-wall data.

The experimental and numerical data are used to study the inclined jet in crossflow. Under the present conditions, the jets always separate from the bottom wall after injection, but the $r=1$ jet re-attaches while the other two do not. There is also some slight but persistent asymmetry, which can be attributed to the plenum feeding mechanism. Finally, the hole in the present jet is short, as is common in film cooling applications; this causes a highly complex flow in the jet as it meets the crossflow, which is usually not accounted for in previous literature. The flow structure in the hole does not depend much on the blowing ratio in the range studied, however. The analysis of the mean flow also serves to validate the simulation data. In general, the agreement is excellent.

Using the MRI data to validate the LES is especially valuable in a complex non-homogeneous flow since data are available in 3D and mean flow structures can be compared. Access to the full field creates confidence that the whole simulated flow is physical, instead of the unfortunate scenario in which a single number (such as velocity/temperature at a single point, or a mean drag/lift coefficient) is matched despite the predicted flow being qualitatively incorrect. Also, if any discrepancies are found it is easier to trace back what aspects of the simulation might be causing the differences. In the present paper, we performed not just qualitative comparisons, but also quantitative ones that leverage the full 3D domain to calculate metrics, such as the circulation analysis. For other flows, it is important to design these metrics such that (i) they represent the flow physics that one desires to capture and (ii) they are robust to MRI noise and possible misalignment (MRI data consist solely of a Cartesian mesh of data, without the exact location of the walls). The latter constraint usually precludes an analysis in which simulation data are interpolated into the MRI mesh and the values are compared cell-by-cell, because this is highly sensitive to misalignment in locations where local gradients are high. Instead, we compared circulations that were computed independently in each dataset, and quantified the uncertainty due to potential misalignment in the MRV; this metric is useful since it captures the strength of the mean counter rotating vortex pair.

In the present work, some turbulence data were presented mostly to demonstrate the enrichment potential. Such data will be further analyzed in future work, especially in regards to the turbulent scalar flux. The datasets presented here consist of a high quality and validated set of simulations with parameter variation (in this case, the velocity ratio $r$) that will be used for data-driven turbulence modeling in film cooling flows, as was done by \citet{milani2018approach}. They can also be made available to any researcher upon request.

\bibliographystyle{model1-num-names}
\bibliography{Milani_IJHFF}

\end{document}